\documentclass[aps,prc,twocolumn,showpacs,showkeys,amsmath,amssymb]{revtex4}
\usepackage{epsfig}
\usepackage{graphicx}% Include figure files
\usepackage{dcolumn}% Align table columns on decimal point
\usepackage{bm}% bold math
\usepackage{mathtext}
\usepackage{graphics}
\usepackage{mathrsfs}
\usepackage{amsmath}
\usepackage{amssymb,latexsym}
\newcommand{\eq}{\begin{eqnarray}}
\newcommand{\en}{\end{eqnarray}}
\newcommand{\bea}{\begin{eqnarray}}
\newcommand{\eea}{\end{eqnarray}}

\newcommand{\ra}{\rangle}
\newcommand{\la}{\langle}

\begin{document}

\title{Role of the $\rho$ meson in the description of\\ 
       pion electroproduction experiments at JLab} 

\author{
Amand Faessler$^1$,
Thomas Gutsche$^1$, 
Valery E. Lyubovitskij$^1$
\footnote{On leave of absence
from Department of Physics, Tomsk State University,
634050 Tomsk, Russia}, 
Igor T. Obukhovsky$^2$ 
\vspace*{1.2\baselineskip}}

\affiliation{$^1$ Institut f\"ur Theoretische Physik,
Universit\"at T\"ubingen,
\\ Auf der Morgenstelle 14, D-72076 T\"ubingen, Germany
\vspace*{1.2\baselineskip} \\
\hspace*{-1cm}$^2$ 
Institute of Nuclear Physics, Moscow
State University,119899 Moscow, Russia 
\vspace*{0.3\baselineskip}\\}

\date{\today}

\begin{abstract}

We study the $p(e,e^\prime\pi^+)n$ reaction in the framework of an
effective Lagrangian approach including nucleon, $\pi$ and $\rho$ meson 
degrees of freedom and show the importance of the $\rho$-meson $t$-pole 
contribution to $\sigma_T$, the transverse part of cross section.
We test two different field representations of the $\rho$ meson, 
vector and tensor, and find that the tensor representation of the
$\rho$ meson is more reliable in the description of the existing data.
In particular, we show that the  $\rho$-meson $t$-pole contribution, 
including the interference with an effective non-local contact term, 
sufficiently improves the description of the recent 
JLab data at invariant mass $W\lesssim$ 2.2 GeV and 
$Q^2\lesssim$ 2.5 GeV$^2/c^2$. A ``soft'' variant of the strong 
$\pi NN$ and $\rho NN$ form factors is also found to be compatible 
with these data. On the basis of the successful description of both the
$\sigma_L$ and $\sigma_T$ parts of the cross section we discuss the importance 
of taking into account the $\sigma_T$ data when extracting the charge pion
form factor $F_\pi$ from $\sigma_L$.

\end{abstract}



\maketitle

%%%%%%%%%%%%%%%%%%%%%%%%%%%%%%%%%%%%%%%%%%%%%%%%%%%%%%%%%%%%%%%%%%%%%%
\section{\label{sec:level1} Introduction.}
%%%%%%%%%%%%%%%%%%%%%%%%%%%%%%%%%%%%%%%%%%%%%%%%%%%%%%%%%%%%%%%%%%%%%%%%

The motivation of many experiments~\cite{Brauel:1979zk,Bebek:1977pe,%
Volmer:2000ek,Huber:2003kg,Horn:2006tm,Tadevosyan:2007yd} 
on forward pion electroproduction at large $Q^2 $ is the study (through 
measurements close to the pion mass shell) of the pion charge form 
factor $F_\pi(Q^2)$. 
For values of the $s$-channel $p+\gamma^*$ energy $W$ above the 
resonance region and
for  small momentum -k ($k^2=t$) transfered to the nucleon spectator the
longitudinal part $\sigma_L$ of the cross section is dominated by the
$t$-channel quasi-elastic mechanism (see Fig.~\ref{fig1}). 
In this case the $\pi NN$ strong form factor 
$F_{\pi{\scriptscriptstyle NN}}^2(t)\approx 1$ 
is a slowly varying function at $|t|\approx M_\pi^2$ and  
\eq
\sigma_L\sim\frac{|t|
F_{\pi{\scriptscriptstyle NN}}^2(t)}{(t-M_\pi^2)^2}
\sigma_{e\pi}^{\rm free}\,, \quad\quad 
\sigma_{e\pi}^{\rm free}\sim F_\pi^2(Q^2) \,, 
\en 
where $\sigma_{e\pi}^{\rm free}$ is the free $e\pi$ cross section. 
However, with currently available data~\cite{Brauel:1979zk,Bebek:1977pe,%
Volmer:2000ek,Huber:2003kg,Horn:2006tm,Tadevosyan:2007yd} 
on the Rosenbluth separation of $\sigma=\sigma_L+\sigma_T$
the situation is not so simple. For comparison of data
to theoretical predictions one should calculate both $\sigma_L$ and
$\sigma_T$ parts of the cross section at least on the basis of a sum
of the $s(u)$- and $t$-pole diagrams depicted in Fig.~\ref{fig2}. 
It should be noted that the $s(u)$-channel contributions 
(Figs.~\ref{fig2}(b) and \ref{fig2}(c)) to the forward pion cross section 
are suppressed~\cite{Speth:1995er} only at considerably high $Q^2$, 
i.e. in the region  
$Q^2\gtrsim$ 2 - 3 GeV$^2$/c$^2$, where the product of corresponding 
vertex form factors and propagators drop faster in
$Q^2$ ($\sim Q^{-n}$, $n\gtrsim$4) as compared to the $\sim Q^{-2}$ 
behavior of the pion form factor $F_\pi(Q^2)$. In this $Q^2$ region the 
forward pion cross section is not as large as for smaller 
$Q^2\lesssim$ 1 GeV$^2$/c$^2$ studied 
earlier~\cite{Brauel:1979zk,Volmer:2000ek}, and until recently
the available data on 
Rosenbluth separation were too poor~\cite{Volmer:2000ek,Huber:2003kg}
to be a reliable basis for the evaluation of $F_\pi(Q^2)$.  

The high quality data recently obtained at 
JLab~\cite{Horn:2006tm,Tadevosyan:2007yd} for $Q^2=$ 1.6 and
2.45 GeV$^2$/c$^2$ can considerably aid in the study of $F_\pi(Q^2)$.
However, at the values $W=$ 1.95 and 2.2  GeV characteristic of the high $Q^2$ 
JLab data (old and new) the kinematical limit for the momentum transfer 
$t_c \approx - (0.1 - 0.15)$  GeV$^2$/c$^2$ is not so close to 
the pion pole position as is the case for low $Q^2\lesssim$ 1 GeV$^2$/c$^2$. 
Thus the vertex dependence on $t$ for all the diagrams in 
Fig.~\ref{fig2} becomes very important for the extraction of 
the pion form factor $F_\pi(Q^2)$ from the data on $\sigma_L$.

In this context the existing data on $\sigma_L$ and $\sigma_T$ available 
now~\cite{Volmer:2000ek,Horn:2006tm,Tadevosyan:2007yd} for a large region 
of momentum transfers 
$0.05 \lesssim - t\lesssim 0.5$ GeV$^2$/c$^2$ can also be used for 
the study of the strong meson-nucleon form factors 
$F_{\pi{\scriptscriptstyle NN}}(t)$,
$G_{\rho{\scriptscriptstyle NN}}(t)$ and $F_{\rho{\scriptscriptstyle NN}}(t)$
in parallel to the study of $F_\pi(Q^2)$.
The direct measurement of these form factors would be very useful for 
both meson-exchange models of the nuclear force~\cite{Machleidt:1987hj} 
and of exchange currents in nuclei~\cite{Riska:1989bh}. Cutoff parameters 
$\Lambda_{{\scriptscriptstyle MNN}}$  
($M  = \pi$ or $\rho$) used in the monopole parametrization 
\eq 
F_{\pi{\scriptscriptstyle NN}}(t) 
=\frac{F_{\pi{\scriptscriptstyle NN}}(0)} 
{1 - t/\Lambda_{\pi{\scriptscriptstyle NN}}^2}\,, \quad
F_{\rho{\scriptscriptstyle NN}}(t) 
=\frac{F_{\rho{\scriptscriptstyle NN}}(0)} 
{1 - t/\Lambda_{\rho{\scriptscriptstyle NN}}^2}  
\label{cut}
\en 
are presently known only indirectly from data and values of 
$\Lambda_{{\scriptscriptstyle MNN}}$ are varying 
in a wide interval of $M_\rho\lesssim \Lambda_{{\scriptscriptstyle MNN}} 
\lesssim 2M_\rho$ (exchange currents in nuclei are usually 
fitted with soft form factors with 
$\Lambda_{{\scriptscriptstyle MNN}} \approx M_\rho$, while 
the nucleon-nucleon interaction models require harder form factors with 
$\Lambda_{{\scriptscriptstyle MNN}} \approx$ (1.5 - 2)$M_\rho$). 
On the other hand, in the constituent quark model 
(CQM)~\cite{LeYaouanc:1972ae,Ackleh:1996yt,Neudatchin:2004pu} 
the cutoff parameter $\Lambda_{{\scriptscriptstyle MNN}}$,
at least for form factors at small values of $- t\lesssim$
0.3 GeV$^2$/c$^2$, is determined by the radius $b$ of the three-quark  
system and for realistic values $b\approx$ 0.5 - 0.6 fm one obtains 
$\Lambda_{{\scriptscriptstyle MNN}}^2 \approx$ 0.5 - 0.7 GeV$^2$/c$^2$.

In Refs.~\cite{Neudatchin:2004pu,Obukhovsky:2005sa} it was shown that the 
recent JLab data on forward pion electroproduction~\cite{Volmer:2000ek} 
are compatible with a soft $\pi NN$ form factor.  
However, in Ref.~\cite{Volmer:2000ek} the data were described on 
the basis of a Regge model modified by introducing a common electromagnetic
form factor $F_\pi(Q^2)$ for both $t$-channel and ``reggeized'' $s$-channel 
amplitudes~\cite{Vanderhaeghen:1997ts} as is schematically shown 
in Fig.~\ref{fig3}. In this model proposed by Vanderhaeghen, Guidal and 
Laget (VGL) there are no constraints on the maximum value of $|t|$ 
as the $t$-dependence of the cross section is determined by the
$t$- (and $s$-) behavior of the $\pi$- and $\rho$-Regge trajectories. 
The model of VGL offers a satisfactory explanation of both
photo- and electroproduction data for pions in a large interval of $t$
including the longitudinal part of the forward pion cross section $\sigma_L$ 
measured at JLab, but fails in explaining the transverse part $\sigma_T$. 
The prediction for $\sigma_T$ is about an order of magnitude smaller than
measured values.
At small $|t|\lesssim$ 0.3 GeV$^2$/c$^2$ a conventional model predicts, 
on the basis of the $t$-pole contribution, the same results starting from the
form factors (\ref{cut}) motivated by the CQM. This was shown in our 
previous work~\cite{Obukhovsky:2005sa} by comparison of predictions 
for $\sigma_L$ made in both models. The situation for $\sigma_T$ is 
also similar (see below), i.e. at small $t$, an approach using strong vertex 
form factors based on the CQM is equally good in the explanation of 
$\sigma_L$, but also fails for $\sigma_T$.

In both models the $\rho$ exchange has little influence on $\sigma_L$ 
for small values of $-t$, while $\sigma_T$ is rather sensitive to this 
contribution.  Hence, a comprehensive analysis of the role of the $\rho$ 
meson for describing the transverse cross section $\sigma_T$ is required. 
Since the introduction and discovery of the vector meson resonances 
their special role was recognized
in phenomena both in nuclear and particle physics~\cite{Sakurai:1967jj}). 
Essential properties of vector 
mesons (e.g. the $\rho$ meson) such as universality and dominance in 
electromagnetic hadronic form factors had a large impact on the 
understanding of the electromagnetic structure of hadrons. Since the early 
sixties attempts to include the vector mesons in the formalism of 
quantum field theory have been initiated~(see \cite{Sakurai:1967jj} 
and references therein) including effective chiral Lagrangian 
approaches~\cite{Schwinger:1967tc}-\cite{Schindler:2005ke}. 
A detailed investigation of different ways to include massive 
vector mesons in the effective low-energy Lagrangians have been 
performed in Ref.~\cite{Ecker:1989yg}. In particular, it was 
shown that the pure tensor representation of vector mesons is most natural 
for constructing their coupling to pseudoscalar mesons and photons 
(the extension onto the baryon sector was done in Ref.~\cite{Borasoy:1995ds})  
consistent with chiral symmetry, vector meson dominance (VMD) and  
asymptotic QCD behavior. However, the conventional vector representation of 
vector mesons is in conflict with VMD and the asymptotic properties of 
QCD~\cite{Ecker:1989yg}. In Ref.~\cite{Ecker:1989yg} for the example of 
the pion electromagnetic form factor, it was demonstrated in the context 
of chiral perturbation theory (ChPT)~\cite{Weinberg:1978kz,Gasser:1983yg} 
how to remedy the shortcomings of the vector representation:
an appropriate local term of order of $O(p^4)$ has to be introduced 
in addition. The corresponding coupling of the local term 
was fixed to achieve a complete agreement between the two schemes based on 
the tensor and vector field representations of the vector mesons. 

Chiral symmetry plays 
an important role in the low-energy domain (below 1 GeV) of quantum 
chromodynamics: it governs the strong interaction between hadrons.
All known low-energy approaches (effective field theories, different
types of quark models, etc.) in the study of the properties of light hadrons
have to incorporate the concept of at least an approximate chiral symmetry
to get reasonable agreement with data. In our case, contrary to what might 
be naively expected for the high energy process $p(e.e^\prime\pi)n$, 
the $t$-channel 
contribution [Fig.~\ref{fig2}(a)] to the quasi-elastic pion knockout
corresponds to a transfer of low energy $k_0=t/2m_{\scriptscriptstyle N}$ and 
momentum $|{\bf k}|= \sqrt{k_0^2 -t}$ to the nucleon spectator, 
and thus a low-energy approach to the $t$-channel terms might well be 
substantiated.

However, in practice we have the interplay of two energy regimes: 
on the one hand the
low-energy dynamics of nucleons with low-momentum $\pi$- and $\rho$-mesons 
in the $t$-channel, 
and on the other hand the initial photon with a large Euclidean 
momentum transfer squared $Q^2 \gtrsim$ 1 - 2 GeV$^2$/c$^2$ and the
final pion with a large energy $E_\pi \gtrsim$ 2 GeV. 
The diagrams contributing to the pion electroproduction in the relevant 
kinematical regime are displayed in Fig.~\ref{fig2}: $t$-channel resonance
diagrams with $\pi$ and $\rho$ exchange in Fig.~\ref{fig2}(a), 
$s$- and $u$-pole diagrams with the intermediate nucleon and a tower of 
nucleon resonances in Fig.~\ref{fig2}(b) and the contact $\gamma\pi NN$ 
diagram (the Kroll-Ruderman term) of Fig.~\ref{fig2}(c). 

On the basis of these Feynman diagrams the main objective of the present work
is to study how the predictions for $\sigma_L$ and $\sigma_T$ depend on the 
$\rho$ meson representation. In our calculations we accept the following 
strategy. 
  
First, we consider the transverse cross section $\sigma_T$ and 
show that it can be well described by taking the $\rho$ meson-exchange 
diagram only. 
The quality of the description is valid up to $Q^2 \simeq 2 $ GeV$^2$.   
To our knowledge this is the first successful description of $\sigma_T$ 
including the special role of the intermediate $\rho$ meson.  
We test both representations of the $\rho$ meson: tensor and vector. 
Our result is that the tensor representation gives a sufficiently better 
description of $\sigma_T$. Of course, both representations can be put 
in equivalence following the idea of Ref.~\cite{Ecker:1989yg} by 
adding an appropriative local term to the Lagrangian of the vector 
representation. We do not resort to this procedure, instead we argue that 
the pure tensor variant is more appropriate from a phenomenological point 
of view and constraints dictated by VMD and the asymptotic QCD behavior. 

Second, we consider the longitudinal cross section $\sigma_L$ and 
show that this quantity asks for a more sophisticated interplay of 
different diagrams from the set of Fig.2. Any description of the $s(u)$-channel
contributions to $\sigma_L$ in terms of nucleon resonances
would be rather complicated and might lead to doubtful results.
Here we follow the results of our recent work~\cite{Obukhovsky:2005sa}. 
On the basis of a quark model it was shown that in the region of 
intermediate $Q^2$ ($\gtrsim$ 1 - 2 GeV$^2$/c$^2$) the effective description 
of $s(u)$-channel and contact-term contributions might be reduced to a 
renormalization of the Kroll-Ruderman contact term modified by strong 
and electromagnetic form factors. The renormalization constant 
is the sole free parameter which we fit to the $\sigma_L$ data.

 Our main finding is that a realistic description of both $\sigma_L$ and
$\sigma_T$ can be obtained in a $t$-channel $\pi+\rho$ approach
with standard values of coupling constants and cutoff parameters,
if: i) we use the tensor representation for the $\rho$ meson leading to
the reproduction of data on the transverse cross section; ii) we approximate
the sum of all the $s$-channel diagrams in Figs.~\ref{fig2}(b) and 
\ref{fig2}(c) by a single effective contact term of type Fig.~\ref{fig2}(c)
with a phenomenological form factor.
For simplicity and to reduce the number of possible free parameters
we choose the cutoff parameter in this form factor to be close to
the one used e.g. in the $\rho\pi\gamma$ form factor. The normalization
of this term is fitted to the $\sigma_L$ data.

In the present manuscript we proceed as follows. First, in Section II
we discuss the basic notions of our approach. We derive the effective
Lagrangian for the description of pion electroproduction off the nucleon.
We discuss different field representations of the $\rho$ meson. 
Then we discuss the contributions of different Born diagrams to  
the amplitude of pion electroproduction. 
In Section III we discuss the choice of hadronic form factors parametrizing 
finite size effects due to hadronic interactions including the photons. 
In Section IV our results are presented in comparison to the JLab data
and to the predictions of the VGL model. 
Finally, in Section V we give a short summary of our results and discuss 
the importance of taking into account the $\sigma_T$ data when extracting 
$F_\pi$ values from the data on $\sigma_L$.

%%%%%%%%%%%%%%%%%%%%%%%%%%%%%%%%%%%%%%%%%%%%%%%%%%%%%%%%%%%%%%%%%%%%%%%%%%%%%%%

\section{Effective Lagrangian and Matrix Elements} 

%%%%%%%%%%%%%%%%%%%%%%%%%%%%%%%%%%%%%%%%%%%%%%%%%%%%%%%%%%%%%%%%%%%%%%%%%%%%%%%

Our considerations for the pion electroproduction are based on an 
effective Lagrangian approach. It involves nucleon, pion, $\rho$ meson 
and photon degrees of freedom. The finite size effects of hadronic 
interactions are parametrized by corresponding form factors. 

\subsection{Inclusion of nucleons, pions and photons} 

The part of the full Lagrangian including the doublet of nucleons $N=(p,n)$, 
the triplet of pions $\vec\pi$ and the electromagnetic field 
$A_\mu$ is motivated by chiral perturbation theory 
(ChPT)~\cite{Weinberg:1978kz,Gasser:1983yg,Gasser:1987rb} and has 
the standard form: 
\eq\label{L_eff} 
{\cal L}_{\rm eff} &=& \bar N ( \not\!\! D - m_N ) N  
+ \frac{1}{2} [ D_\mu\vec\pi D^\mu\vec\pi - M_\pi^2 \vec\pi^2 ] 
\nonumber\\ 
&-& \frac{g_A}{2F_\pi} \bar N D_\mu\vec\pi \vec\tau \gamma^\mu\gamma^5 N   
- \frac{1}{4} F_{\mu\nu} F^{\mu\nu} + \cdots 
\en 
where $\pi^\pm = - (\pi_1 \mp i \pi_2)/\sqrt{2}$, $\pi^0 = \pi_3$, 
$F_{\mu\nu} = \partial_\mu A_\nu - \partial_\nu A_\mu$ is the stress 
tensor of the electromagnetic field,  $g_A$ is the nucleon axial 
charge, $F_\pi$ is the leptonic decay constant, 
$m_N \equiv m_p = 938.27$~MeV and $M_\pi \equiv M_{\pi^\pm} = 139.57$~MeV 
are nucleon and pion masses. The symbol $\cdots$ denotes terms of 
higher order not needed in our consideration. In the numerical calculations 
we express $g_A/(2F_\pi) = g_{\pi NN}/(2m_N)$ through the strong $\pi NN$ 
coupling constant using the Goldberger-Treiman relation with 
$g_{\pi NN} = 13.5$. The covariant derivative $D_\mu$, containing the
electromagnetic field and acting on proton and charged pion fields, 
is defined as: 
$D_\mu p = (\partial_\mu - ieA_\mu) p\,$ and 
$D_\mu \pi^\pm = (\partial_\mu \mp ieA_\mu) \pi^\pm\,,$ 
where $e$ is the proton charge. For neutral fields (neutron and $\pi^0$) 
$D_\mu$ coincides with ordinary derivative. The inclusion of the $\rho$ 
meson and the addition of strong and electromagnetic form factors in the 
effective Lagrangian (\ref{L_eff}) will be discussed below. Note, in 
addition to the more convenient pseudovector (PV) coupling of the pion to 
nucleons (the third term in r.h.s. of Eq.~(\ref{L_eff})) we also consider 
the pseudoscalar (PS) coupling: 
${\cal L}_{\pi NN}^{PS} \, = \, 
g_{\pi NN} \, \bar N \, i \gamma^5 \, \vec\pi\vec\tau \, N \,.$ 

\subsection{Inclusion of vector mesons} 
\label{sec:currents}

%%%%%%%%%%%%%%%%%%%%%%%%%%%%%%%%%%%%%%%%%%%%%%%%%%%%%%%%%%%%%%%%%%%%%%%%%%%%%%%

For the $\rho$ meson we use two different field representation: tensor and 
vector. Here we follow Refs.~\cite{Gasser:1983yg,Ecker:1988te,Ecker:1989yg,%
Borasoy:1995ds,Kubis:2000zd}. 
In the tensor representation the triplet of $\rho$ mesons  
is written in terms of antisymmetric tensor fields: 
$\rho_{\mu\nu}^W = - \rho_{\nu\mu}^W 
= ( \vec{\rho}^{\, W} \, \vec{\tau} )_{\mu\nu}/\sqrt{2} $.   
The free Lagrangian of vector mesons in the tensor representation is written 
in the form 
\eq 
{\cal L}_{\rho}^W = -\frac{1}{2}\partial^\mu \rho_{\mu\nu}^{W a} 
\partial_\alpha \rho^{W \alpha\nu}_a
+\frac{M_\rho^2}{4}\rho_{\mu\nu}^{W a} \rho^{W \mu\nu}_a.
\label{lagrw}
\en 
The $\rho$-meson propagator in the tensor representation 
has the form 
\begin{multline} 
G_{W; \mu\nu,\alpha\beta}^{ab}(x-y)=
\langle0|T\{\rho_{\mu\nu}^{W a}(x),\rho_{\alpha\beta}^{W b}(y)\}|0\rangle\\
= - \frac{\delta^{ab}}{M_\rho^2}\int\frac{d^4k}{(2\pi)^4 i}
\frac{e^{ik\cdot(x-y)}}{M_\rho^2-k^2-i\epsilon} 
[g_{\mu\alpha}g_{\nu\beta}(M_\rho^2-k^2)\\
+g_{\mu\alpha}k_{\nu}k_{\beta}
-g_{\mu\beta}k_{\nu}k_{\alpha}-(\mu\leftrightarrow\nu)]\,, 
\label{propw}
\end{multline}
where the terms in the square brackets proportional to $(M_\rho^2-k^2)$ 
will generate contact terms in the vector-meson exchange interaction. 

Now we turn to a discussion of the vector representation of $\rho$ mesons, 
i.e. in terms of the vector fields $\rho_\mu$.  
The corresponding free Lagrangian has the form: 
\eq 
{\cal L}_\rho^V =-\frac{1}{4}\rho_{\mu\nu}^{V a} \rho^{V a \mu\nu} 
+\frac{M_V^2}{2} \rho_\mu^{a} \rho^{a \mu},
\en
where $\rho_{\mu\nu}^{V a} = \partial_\mu \rho_\nu^{a} 
- \partial_\nu \rho_\mu^{a}$. 
For the sake of comparison between the two different representation it is 
convenient to write down the propagator in the vector representation 
as a $T$-product of $\rho^V_{\mu\nu}$: 
\eq 
& &G_{V; \mu\nu,\alpha\beta}^{ab}(x-y)=
\langle0|T\{\rho^{V a}_{\mu\nu}(x),\rho^{V b}_{\alpha\beta}(y)\}
|0\rangle\nonumber\\
&=& - \frac{\delta^{ab}}{M_V^2}\int\frac{d^4k}{(2\pi)^4 i}
\frac{e^{ik\cdot(x-y)}}{M_V^2-k^2-i\epsilon}\nonumber\\
&\times&[g_{\mu\alpha}k_{\nu}k_{\beta}
-g_{\mu\beta}k_{\nu}k_{\alpha}-(\mu\leftrightarrow\nu)] \,.
\label{propv2} 
\en 
As was stressed in Ref.~\cite{Ecker:1989yg} the propagators 
$G_{W; \mu\nu,\alpha\beta}^{ab}$ and $G_{V; \mu\nu,\alpha\beta}^{ab}$ 
differ by the contact term contained in the tensorial propagator: 
\begin{multline} 
G_{W; \mu\nu,\alpha\beta}^{ab}(x) = 
G_{V; \mu\nu,\alpha\beta}^{ab}(x)\\ 
+ \frac{i}{M_V^2}[g_{\mu\alpha}g_{\nu\beta} - g_{\mu\beta}g_{\nu\alpha}] \, 
 \delta^{ab} \, \delta^4(x) \,. 
\end{multline}  
Therefore, the use of the two different representations le\-ads to a 
different off-shell behavior of vector meson-exchange diagrams. 
As we already mentioned in Introduction, a detailed analysis of 
vector and tensor schemes was performed in Ref.~\cite{Ecker:1989yg} 
for the example of the electromagnetic pion form factor. The 
tensor representation was found to be fully consistent with 
constraints of chiral symmetry, VMD and the asymptotic behavior of QCD. 
To get equivalence of the two representations a certain inclusion of 
an additional local term was required. In our analysis we find 
that the tensor representation for the vector mesons is more reliable 
and leads to an understanding of the transverse cross section of the pion 
electroproduction in the considered kinematical situation. 
In particular, the additional contact term in the propagator of the tensor 
representation considerably modifies the $\rho$-exchange 
contribution to the cross section.  

According to~\cite{Borasoy:1995ds,Kubis:2000zd}, a chirally invariant 
Lagrangian for the couplings of the tensor field $\rho^W_{\mu\nu}$ to 
baryons can be written in the general form containing couplings which 
can be related to the ones of the commonly used vector representation. 
Therefore, we further proceed using the vector representation 
for the $\rho NN$ couplings:  
\eq 
{\cal L}_{\rho NN}=\frac{1}{2}
\bar N\left(G_{\rho NN} \vec\rho_\mu \gamma^\mu
-\frac{F_{\rho NN}}{2m_{\scriptscriptstyle N}}\sigma^{\mu\nu}\partial_\nu 
\vec\rho_\mu \right) \, \vec\tau \, N \,. 
\label{lvn}
\en 
The anomalous $\rho\pi\gamma$ coupling is defined as 
\eq 
{\cal L}^{V(W)}_{\rho\pi\gamma} \, = \, \frac{e M_\rho}{4} \, 
g_{\rho\pi\gamma} \, \varepsilon^{\mu\nu\alpha\beta} \, F_{\mu\nu} \, 
\vec\rho_{\alpha\beta}^{\, V(W)}  \, \vec\pi \,. \label{rpgvw} 
\en 
The coupling constant $g_{\rho\pi\gamma} = 0.728$ GeV$^{-1}$ is 
fixed from the $\rho\to\pi\gamma$ decay width: 
\eq 
\Gamma(\rho \to \pi\gamma) = \frac{\alpha}{24} \, 
g_{\rho\pi\gamma}^2 \, M_\rho^3 \, 
\biggl[ 1 - \frac{M_\pi^2}{M_\rho^2}\biggr]^3 \,, 
\en 
where $\alpha = 1/137.036$ is the fine-structure constant. 
In our convention the isospin symmetric hadron masses of the iso-multiplets
are identified with the masses of the charged partners:
\eq 
\hspace*{-.5cm}
& &m_{\scriptscriptstyle N}=m_p=938.27 \ {\rm MeV}\,, \ \ 
M_{\pi}=M_{\pi^\pm}=139.57 \ {\rm MeV}\,,\nonumber\\
\hspace*{-.5cm}
& &M_{\rho}=M_{\rho^\pm}=775.5 \ {\rm MeV}\,.
\en

\subsection{Born diagrams contributing to the pion electroproduction }

In the calculation of the amplitude for the pion electroproduction off the
nucleon we restrict to the Born approximation. At the order of accuracy 
we are working in we include t-channel diagrams with $\rho$ and 
$\pi$ exchange [Fig.2(a)], $s$- and $u$-channel diagrams with 
intermediate nucleons [Fig.2(b)] and in the case of the pseudovector 
coupling of pions to nucleons we have the extra diagram [Fig.2(c)], 
the so-called Kroll-Ruderman term, describing the contact 
coupling of the photon to two nucleons and one pion. 

\subsubsection{Contribution of the $\rho$-meson exchange diagram }

We start with the discussion of the $\rho$-meson exchange diagram. 
Despite the difference between the propagators and between the 
$\rho\pi\gamma$ couplings in the respective representations
the final expression has a common universal form.
For the $\rho$-meson  exchange diagram contribution to the pion 
electroproduction amplitude [Fig.~\ref{fig2}(a)] for both the $W$ and $V$ 
variants we have
\begin{widetext}
\eq 
T_\rho^\bigg{\{\!\!\begin{array}{c}
{\scriptstyle V}\\[-4pt]
{\scriptstyle W}
\end{array}\!\!\bigg\}}(\lambda,s,s^\prime)&=&
\frac{e}{2\sqrt{2}} \, g_{\rho\pi\gamma} 
\varepsilon_{\mu\nu}^{(\lambda)}(q)\bar u_n(p^\prime,s^\prime)
\biggl\{\biggl[G_{\rho{\scriptscriptstyle NN}}+
\frac{\bigg\{\!\!\begin{array}{c}
{\scriptstyle k^2}\\[-4pt]
{\scriptstyle M_\rho^2}
\end{array}\!\!\bigg\}}{4m_{\scriptscriptstyle N}^2}
F_{\rho{\scriptscriptstyle NN}}\biggr]
\frac{2m_{\scriptscriptstyle N}}{M_\rho^2\!-\!k^2}\sigma^{\mu\nu}
\gamma_5  \nonumber\\
&-&\frac{2m_{\scriptscriptstyle N}}{M_\rho^2\!-\!k^2}
\biggl[G_{\rho{\scriptscriptstyle NN}}\frac{P^\mu\gamma^\nu\!-
\!P^\nu\gamma^\mu}{2m_{\scriptscriptstyle N}}
-F_{\rho{\scriptscriptstyle NN}}\frac{P^\mu k^\nu\!-
\!P^\nu k^\mu}{4m_{\scriptscriptstyle N}^2}
\biggr]i\gamma_5\biggr\}u_p(p,s)\,, 
\label{rho}
\en 
\end{widetext}
where $P=p+p^\prime$, $k=p-p^\prime$ and $s,s^\prime$ 
denote the spin projections of the initial proton and final neutron, 
respectively; $\varepsilon_{\mu\nu}^{(\lambda)}(q) = 
q_\mu\epsilon^{(\lambda)}_\nu(q) - q_\nu\,\epsilon^{(\lambda)}_\mu(q)$. 
Here $\epsilon_\mu^{(\lambda)}(q)$, with $\lambda{=}0,\pm 1$, are basis 
vectors of circular polarization for the virtual photon quantized along 
the momentum ${\bf q}$, i.e. they are defined as 
\begin{multline} 
\epsilon^{(\lambda\!=0)\mu}(q)=
\biggl\{\frac{|{\bf q}|}{Q},0,0,\frac{q_0}{Q}\biggr\},\\
\epsilon^{(\lambda\!=\pm 1)\mu}(q)=
\biggl\{0,\mp\frac{1}{\sqrt{2}},\frac{-i}{\sqrt{2}},0\biggr\},\quad 
Q=\sqrt{-q^2}\,. 
\end{multline} 
The vectors $\epsilon_\mu^{(\lambda)}(q)$ satisfy the 
conventional orthogonality, normalization and completeness 
conditions~\cite{Budnev:1975de}
\eq 
q^\mu\epsilon_\mu^{(\lambda)}(q)=0\,,\quad
\epsilon^{(\lambda)\mu}(q){\epsilon^{(\lambda^\prime)}_\mu}^\ast(q)
=(-1)^\lambda\delta^{\lambda\lambda^\prime},\nonumber\\
\sum_{\lambda\!=0,\pm 1}(-1)^\lambda\epsilon^{(\lambda)}_\mu(q)
{\epsilon^{(\lambda)}_\nu}^\ast(q)=g_{\mu\nu}-\frac{q_\mu q_\nu}{q^2} \,. 
\label{proj}
\en 
In Eq.~(\ref{rho}) the factor 
$\{\!\!\begin{array}{c}
{\scriptstyle k^2}\\[-4pt]
{\scriptstyle M_\rho^2}
\end{array}\!\!\}$ in the first square 
brackets should be taken equal to $k^2$ for the vector $V$-variant 
and equal to $M_\rho^2$ for the tensor $W$-variant. From Eq.~(\ref{rho})  
one can conclude that the results for the $W$ and $V$ variants are degenerate 
when $k^2 \to M_\rho^2$, but in the region $(M_\rho^2-k^2)\approx m_N^2$ 
characteristic of the JLab data~\cite{Volmer:2000ek,Huber:2003kg,%
Horn:2006tm,Tadevosyan:2007yd} they are considerably different. 
The corresponding $\rho$-induced $\gamma\pi NN$ ``contact'' interaction 
arising in the tensor variant can be defined as
\eq 
T_\rho^{W} \, - \, T_\rho^{V} &=& 
\frac{e}{4 m_{\scriptscriptstyle N} \sqrt{2}} 
\, g_{\rho\pi\gamma} \, F_{\rho{\scriptscriptstyle NN}} \, 
\epsilon_{\mu\nu}^{(\lambda)}(q) 
\nonumber\\
&\times&\bar u_n(p^\prime,s^\prime) \sigma^{\mu\nu} \gamma_5 u_p(p,s)\,.
\label{cont}
\en 
It should be noted that the difference encoded in the pion electroproduction
amplitude in the contact term (\ref{cont}) is sufficient to get a good 
description of the transverse cross section $\sigma_T$. 

\subsubsection{Contribution of the $\pi$-meson exchange diagram }

The pion $t$-pole diagram [Fig.2(a)] gives the following contribution 
to the amplitude 
\eq 
T_\pi(\lambda,s,s^\prime)&=&
e\sqrt{2}g_{\pi\scriptscriptstyle NN}
\frac{\epsilon^{(\lambda)}(q) \cdot (k+k^{\,\prime})}{M_\pi^2\!-\!k^2}
\nonumber\\ 
&\times&\bar u_n(p^{\,\prime},s^\prime) i\gamma_5 u_p(p,s) \,, 
\label{pi} 
\en 
where $k = p - p^{\,\prime}\,, \ $ $t=k^2\,, \ $  
$k^{\,\prime} = k + q \,, \ $ ${k^\prime}^2=M_\pi^2$ and 
$q^2 = - Q^2$.

\subsubsection{Nucleon $s$- and $u$-pole diagrams}

The sum of the nucleon $s$- and $u$-pole diagram contributions to the total 
amplitude is   
\begin{widetext}
\eq 
T_{{\scriptscriptstyle N}(s\!+\!u)}(\lambda,s,s^\prime)&=& 
-\,e \sqrt{2} g_{\pi\scriptscriptstyle NN}
\epsilon^{(\lambda)}_\mu(q)
\bar u(p^\prime,s^\prime)
\biggl[
\biggl\{\!\!\!\begin{array}{c}
i\gamma_5\\[1mm] 
\frac{i \not k^{\,\prime} \gamma_5}{2m_{\scriptscriptstyle N}}
\end{array}\!\!\!\biggr\}
\frac{\not\! p+\!\not\! q\!+\! m_{\scriptscriptstyle N}}
{W^2-m_{\scriptscriptstyle N}^2}\biggl(F_{1p}\gamma^\mu+F_{2p}
\frac{i\sigma^{\mu\nu}q_\nu}{2m_{\scriptscriptstyle N}}\biggr)\nonumber\\
&-&\biggl(F_{1n}\gamma^\mu+F_{2n}
\frac{i\sigma^{\mu\nu}q_\nu}{2m_{\scriptscriptstyle N}}\biggr)
\frac{\not\! p-\!\not\! k^\prime\!+\!m_{\scriptscriptstyle N}}
{W^2\!+\!Q^2+t-M_\pi^2\!-\!m_{\scriptscriptstyle N}^2} 
\biggl[
\biggl\{\!\!\!\begin{array}{c}
i\gamma_5\\[1mm] 
\frac{i \not k^{\,\prime} \gamma_5}{2m_{\scriptscriptstyle N}}
\end{array}\!\!\!\biggr\}
\biggr]u(p,s)\,,  \label{nsu}
\en 
\end{widetext} 
where $s=(p+q)^2=W^2$ and 
$u=(q-p^\prime)^2=-W^2-Q^2-t+2m_{\scriptscriptstyle N}^2+M_\pi^2$. 
Here the factors $i\gamma_5$ or 
$\frac{\not k i\gamma_5}{2m_{\scriptscriptstyle N}}$ in the 
column $\{ \cdots \}$ correspond to the pseudoscalar (PS) or 
pseudovector (PV) $\pi NN$ coupling respectively. 
The coupling constants $F_{iN}$ equal 
\eq 
F_{1p} = 1\,,  \hspace*{.3cm} 
F_{1n} = 0\,,  \hspace*{.3cm} 
F_{2p} = \mu_p - 1\,,  \hspace*{.3cm} F_{2n} = \mu_n\,,  
\en 
where $\mu_p$ and $\mu_n$ are magnetic moments of proton and neutron. 

\subsubsection{The $\gamma \pi NN$ contact diagram} 

The $\gamma \pi NN$ contact diagram of Fig.2(c) only shows up for 
the case of the PV variant.
The corresponding amplitude is (we denote it by the 
subscript $CPV$, that is contact pseudovector coupling): 
\eq 
T_{\scriptscriptstyle CPV}(\lambda,s,s^\prime)&=&
-\,\frac{e\sqrt{2}g_{\pi{\scriptscriptstyle NN}}}{2m_{\scriptscriptstyle N}}
\nonumber\\ 
&\times& \bar u(p^\prime,s^\prime)\not\!\epsilon^{\, (\lambda)}(q) \, 
i\gamma_5 \, u(p,s) \,. 
\label{pvcont}
\en

Finally, we make a comment concerning the interference of the
$\rho$-exchange amplitude with other contributions in the calculation 
of the cross section. In the vector variant the interference term between 
the $T_\pi$ - and $T_\rho^V$ - pole diagrams does not contribute to the 
$\sigma_L$ and $\sigma_T$ cross sections. However, in the tensor
variant the interference term between the $T_\pi$ - and $T_\rho^W$ - pole 
diagrams is not negligible (see below) and thus we should consider
the interference terms for all the potentially important diagrams including 
the $s(u)$-channel diagrams [Fig.~\ref{fig2}(b)].

\section{Form factors}
\label{sec:formf} 

\subsection{General consideration}

Up to now we deal with diagrams generated by the effective Lagrangian 
involving nucleons, pions, the $\rho$ meson and the photon (see discussion in 
previous section). It can be easily checked that the sum of Born 
diagrams (Fig.2) is gauge invariant. 
E.g. the hadronic electromagnetic current 
$\la p^\prime s^\prime|J^\mu_{\rm Born}|ps\ra$ defined as 
\begin{multline} 
\epsilon^{(\lambda)}_\mu 
\langle p^\prime s^\prime|J^\mu_{\rm Born}|ps\rangle
= T_\rho(\lambda,s,s^\prime) + T_\pi(\lambda,s,s^\prime)\\ 
+ T_{n(s+u)}(\lambda,s,s^\prime)
+ T_{\scriptscriptstyle CPV}(\lambda,s,s^\prime) \, 
\label{born}
\end{multline} 
satisfies current conservation 
$q_\mu \la p^\prime s^\prime|J^\mu_{\rm Born}|ps\ra  = 0. \ \ $  
Note, that the $\rho$-meson exchange and pion-pole diagrams satisfy 
current conservation separately, while $s(u)$-pole and contact 
Kroll-Ruderman term are separately do not satisfy this condition 
(i.e. only in sum). 

Now we are in the position to modify the vertices describing strong 
and electromagnetic interactions of hadrons - by introducing 
hadronic form factors. 
The idea of such a modification is clear: we would like to include 
finite size effects. 
The introduction of form factors into the interaction Lagrangian and, 
therefore, into the Born amplitudes leads to a violation 
of gauge invariance. To restore gauge invariance one can use different 
methods (see e.g. discussion in 
Refs.~\cite{Gross:1987bu}-\cite{Anikin:1995cf}). 
One of the methods is based on the Gross-Riska procedure~\cite{Gross:1987bu}, 
which does the following modification in matrix 
elements: every term containing form factor $F$ which is multiplied with 
$\gamma$-matrix and vector $P^\mu$ with open Lorentz index $\mu$ coinciding 
with the index of the photon polarization vector 
is modified as~\cite{Gross:1987bu}: 
\eq\label{Gross_Riska} 
\gamma^\mu F \to \gamma^\mu F + q^\mu \frac{\not\! q}{q^2} ( 1 - F ) 
= \gamma^\mu_\perp  F + q^\mu \frac{\not\! q}{q^2} \,, \nonumber\\
P^\mu F \to P^\mu F + q^\mu \frac{Pq}{q^2} ( 1 - F ) = 
P^\mu_\perp F + q^\mu \frac{Pq}{q^2} \,, 
\en  
where $\gamma^\mu_\perp = \gamma^\mu - q^\mu \!\! \not\!\!\! q/q^2$ 
  and $P^\mu_\perp = P^\mu - q^\mu Pq/q^2$ are the Dirac matrix 
and momentum which are orthogonal to photon momentum and are 
obtained from original quantities by multiplying 
with the projector $g_{\mu\nu}^\perp = g_{\mu\nu} 
- q_\mu q_\nu/q^2$. 
Note that idea suggested in~\cite{Gross:1987bu} was extended to 
pion electroproduction in~\cite{Nozawa:1989pu} and was extensively 
used in Refs.~\cite{Scherer:1991cy,Bos:1992qs}. In particular, 
the Gross-Riska procedure leads to the correct low-energy theorems and 
guarantees that the partial conservation of axial current (PCAC) 
constraint for the pion electroproduction amplitude is 
satisfied~\cite{Scherer:1991cy}. 

In this paper we use the similar method of restoring electromagnetic 
gauge invariance which is fully equivalent to the Gross-Riska 
prescription~\cite{Gross:1987bu} 
when we fulfil the additional conditions (\ref{proj}),
i.e. use the circular polarization for the virtual photon field. 
In particular, our modification of matrix elements reads as: 
\eq\label{our_method} 
\gamma^\mu F \to \gamma^\mu_\perp  F \,, \nonumber\\
P^\mu F \to P^\mu_\perp F \,. 
\en  
An advantage of our method is that each diagram is separately 
satisfy the current conservation by construction, 
due to $q_\mu \gamma^\mu_\perp = 0$ and $q_\mu P^\mu_\perp$. 
It is sufficient in our consideration while instead of sum of the 
$s(u)$-pole and local Kroll-Ruderman term we will use the 
modified Kroll-Ruderman term with form factor (see discussion below), 
which should satisfy the current conservation separately. 
   
When introducing form factors, in the fit to data we intend to deal with a
minimal amount of free parameters which should be common 
for both variants of the $\rho$ representation. For this purpose
we use a common form factor of a simple monopole 
form (\ref{cut}) for all the strong meson-nucleon vertices 
with the same cutoff parameter $\Lambda_{\rm str}$:
\eq 
& &g_{\pi NN}  \to g_{\pi NN}(t) = g_{\pi NN} F_{\rm str}(t)\,, 
\nonumber\\
& &G_{\rho NN} \to G_{\rho NN}(t) = G_{\rho NN} F_{\rm str}(t)\,, \nonumber \\
& &F_{\rho NN} \to F_{\rho NN}(t) = F _{\rho NN} F_{\rm str}(t)\,, \nonumber \\
& &F_{\rm str}(t) = \frac{1}{1 - t/\Lambda^2_{\rm str}} \,. 
\label{lams}
\en 
We vary the parameter $\Lambda_{\rm str}^2$ in the region 
0.5 - 0.7 GeV$^2$/c$^2$ (which is close to the CQM predictions) 
to fit the JLab data on $\sigma_L$.

The form factors for the electromagnetic vertices are known with better 
accuracy, both for the pion and the nucleon. We use a monopole 
parametrization for the pion
\begin{equation}
e \to e F_\pi(Q^2)\,, \hspace*{.5cm}  
F_\pi(Q^2) = \frac{1}{1+Q^2/\Lambda_\pi^2},
\label{lame}
\end{equation}
where $\Lambda_\pi^2$ should be close to its mean value of
$\Lambda_\pi^2\approx$ 0.54 GeV$^2$/c$^2$,
and the dipole parametrization for the electromagnetic Sachs form 
factors of nucleons. 

The form factor of the $\rho\pi\gamma$ vertex is the most uncertain since at
this vertex two variables, $t$ and $Q^2$, are off-shell (for the
$\pi\pi\gamma$ vertex, where the situation is similar, we neglect
the $t$ dependence, since for the forward pion electroproduction 
$t$, as a rule, is close to its on-mass shell value of $t=M_\pi^2$). 
For reasons motivated
by the CQM we modify the $\rho\pi\gamma$ vertex as 
\eq 
g_{\rho\pi\gamma} \to  g_{\rho\pi\gamma}(t,Q^2) =  
g_{\rho\pi\gamma} F_{\rho\pi\gamma}(t,Q^2) 
\en 
where for the form factor $F_{\rho\pi\gamma}(t,Q^2)$ we take the combined
expression
\eq 
F_{\rho\pi\gamma}(t,Q^2)=\frac{1}{1+(M_\rho^2-t)/(4M_\rho^2)}\,
\frac{1}{1+Q^2/\Lambda_{\rm eff}^2}.  
\label{lamr}
\en 
We consider only two possibilities for the $Q^2$ behavior: 
\begin{itemize}
\item[(a)] $\Lambda_{\rm eff}=\Lambda_\pi$, the ``soft'' variant,
\item[(b)] $\Lambda_{\rm eff}\gtrsim 2\Lambda_\pi$, the ``hard'' variant.
\end{itemize}
In the ``hard'' variant $\Lambda_{\rm eff}$ is considered as 
a free parameter close to the usual values of 
$\Lambda_\rho\approx$ 1 - 1.2 GeV/c \cite{Vanderhaeghen:1997ts,Horn:2006tm}. 
We will fit $\Lambda_{\rm eff}$ to the JLab data on $\sigma_T$.

In conclusion we shortly formulate/repeat a common rule for all the vertices
where a pion is created or annihilated. In the Born expression for such a
vertex 
\begin{itemize}
\item[(i)] $eF_\pi(Q^2)$ should be substituted for the charge $e$,
\item[(ii)]$g_{\pi{\scriptscriptstyle NN}}(t)$  
should be substituted for $g_{\pi{\scriptscriptstyle NN}}$.    
\end{itemize} 
This rule is extended to the contact term (\ref{pvcont}) as well (see below).
The modification of the $s$- and $u$-channel contributions including the
contact (Kroll-Ruderman) term will be discussed in the next subsection. 

Some remarks on the $t$-dependence of Eq.~(\ref{lamr}) should be added. 
We use a ``hard'' cutoff parameter 2$M_\rho$ for the $t$-dependence of the
$\rho\pi\gamma$ vertex in both the time-like $t>$ 0 and space-like 
$t<$ 0 regions. In the space-like region the strong hadron form factors have 
been evaluated in many works (see e.g.~\cite{Machleidt:1987hj,Riska:1989bh})
on the basis of a rich data base on $NN$ scattering and exchange currents in 
nuclei,
in which case the value of the cutoff parameter varies between $M_\rho$ and 
2$M_\rho$. But our task is to evaluate the form factor in the time-like region
on the basis of the $\sigma_T$ data. Our efforts to describe $\sigma_T$ 
with the soft cutoff parameter $\sim M_\rho$ fail since in this case the 
effective value of the $\rho\pi\gamma$ coupling in the region near $t\sim$ 0 
is suppressed by the factor $[1+(M_\rho^2-t)/M_\rho^2]^{- 1} \sim 1/2$. 
This probably indicates that in the time-like region the cutoff parameter 
should be hard, and thus here we use the large value 2$M_\rho$. 
We further do not 
vary this parameter to simplify handling other free parameters when fitting 
the cross section.

\subsection{An effective description of ${\bf s}$- and  ${\bf u}$-channel 
contributions}

The only exclusion from the rules $(i)$ - $(ii)$ is the $\pi NN$ vertex in 
the nucleon s- and u-pole diagrams [Fig.~\ref{fig2}(b)], 
where the pion is on its mass-shell
(${k^\prime}^2=M_\pi^2$), but the intermediate nucleon, after absorption of 
a large $Q^2\gtrsim m_{\scriptscriptstyle N}^2$ is severely off its mass 
shell. For this vertex we formally introduce a form factor 
$F_{\rm eff}(t,Q^2)$, but do not really use it in calculations since such 
a form factor should include contributions of all the excited baryon states 
compatible with a large virtual mass $W$ in the $s$-channel. 

Since a description in terms of baryon
poles would be very complicated and in practice is beyond reach we turn
from the hadron picture of the $s$-channel process to the quark model
consideration following our recent work~\cite{Obukhovsky:2005sa}.
Such a consideration gives at least qualitative  insight into the relevant
processes when a large $Q^2$, induced by electroproduction of pions, is
propagating through the three-quark system (Fig.~\ref{fig4}). In the quark
models three mechanisms are implied. The first one [Fig.~\ref{fig4}(a)]  
corresponds to the $t$-channel hadron mechanism considered above with a
small momentum transfer to the nucleon spectator. The other two mechanisms
[Figs.~\ref{fig4}(b) and \ref{fig4}(c)] generate amplitudes which differ 
in the power $n$ of the large $Q^2$ behavior ($\sim Q^{-n}$). 
For the diagram of  
Fig.~\ref{fig4}(c) the amplitude has asymptotic behavior with $n\gtrsim$ 4,
while for the other one [Fig.~\ref{fig4}(b)] $n$ should be smaller and similar 
to the one of the pion form factor with $n=$ 2.

Starting from this observation we discussed the above quark mechanisms 
in Ref.~\cite{Obukhovsky:2005sa} in terms of a naive 
$^3P_0$-model~\cite{LeYaouanc:1972ae,Ackleh:1996yt} 
on the basis of a harmonic oscillator quark model. Our evaluation had shown
that in this approximation the  corresponding effective amplitude 
$T_{\rm eff}$ becomes  proportional to the contact term  (\ref{pvcont}) 
times the product of two form factors, the electric pion and strong $\pi NN$, 
i.e. $T_{\rm eff}\sim T_{\scriptscriptstyle CPV}
F_\pi(Q^2)F_{\pi{\scriptscriptstyle NN}}(t)$ (see 
Ref.~\cite{Obukhovsky:2005sa} for detail). The resulting amplitude is
perfectly in line with the above formulated empirical rules 
$(i)$ and $(ii)$. 

However, at large $Q^2$ the $^3P_0$-model cannot be reliable in predictions 
for the $Q^2$ behavior of the amplitudes and thus, instead of the pion 
form factor $F_\pi(Q^2)$, we use the more general phenomenological form factor
of the form
\eq
F_{\rm eff}(Q^2)=\frac{1}{1+Q^2/\Lambda_{\rm eff}^2}.
\label{feff}
\en
For simplicity we use the same cutoff parameter $\Lambda_{\rm eff}$ 
for all the effective $Q^2$-dependent terms (see, for example, the
analogous effective $\rho\pi\gamma$ form factor of Eq.~(\ref{lamr})) .
 
From the above discussion follows that the effective description
of $s(u)$-channel and contact-term contributions to the pion 
electroproduction amplitude at intermediate values of $Q^2$ might be reduced 
to a renormalization of the contact term modified by
electric $F_{\rm eff}$ and strong $F_{\rm str}$ form factors
\begin{equation}
\left\{T_{{\scriptscriptstyle N}(s\!+\!u)}+T_{\scriptscriptstyle CPV} 
\right\}_{\rm eff}\approx 
Z F_{\rm eff}(Q^2)F_{\rm str}(t)T_{\scriptscriptstyle CPV}.
\label{eff}
\end{equation}
In a simplified model here we consider this possibility by fitting the free 
parameter $Z$ to the data on $\sigma_L$. Hence, the total amplitude is
written in the form
\begin{equation}
T=T_\pi+T_\rho 
+ZF_{\rm eff}(Q^2)F_{\rm str}(t)
T_{\scriptscriptstyle CPV},
\label{simple}
\end{equation}
where $T_{\scriptscriptstyle CPV}$ is the Born amplitude (\ref{pvcont}),
while $T_\rho$ and $T_\pi$ are the amplitudes (\ref{rho}) and (\ref{pi}) 
modified by strong and electromagnetic form factors.

%%%%%%%%%%%%%%%%%%%%%%%%%%%%%%%%%%%%%%%%%%%%%%%%%%%%%%%%%%%%%%%%%%%%%%%%%%%%%%%

\section{Results}
\label{sec:results}

%%%%%%%%%%%%%%%%%%%%%%%%%%%%%%%%%%%%%%%%%%%%%%%%%%%%%%%%%%%%%%%%%%%%%%%

In this section we discuss result of our calculation of 
the transverse and longitudinal cross sections. 

The differential cross section for the $p(e,e^\prime\pi^+)n$ reaction
integrated over the azimuthal angle $\phi_{e^\prime}$ of the electron
in the one-photon approximation is usually defined~\cite{revews} as
\begin{multline} 
\int_0^{2\pi}\frac{d^5\sigma}{dE_{e^\prime}d\Omega_{e^\prime}
d\Omega_{\pi^\prime}^\ast}d\phi_{e^\prime}
\, = \, 2 |{\bf q}^\ast||{{\bf k}^\prime}^\ast|\,2\pi\Gamma_t
\biggl\{\varepsilon\frac{d\sigma_L}{dtd\phi_{\pi^\prime}^\ast}\\
+\frac{d\sigma_T}{dtd\phi_{\pi^\prime}^\ast}
+\varepsilon\frac{d\sigma_{TT}}{dtd\phi_{\pi^\prime}^\ast}
+\sqrt{2\varepsilon(1+\varepsilon)}
\frac{d\sigma_{LT}}{dtd\phi_{\pi^\prime}^\ast}
\biggr\},
\label{sig5}
\end{multline} 
where $\varepsilon$ is the invariant parameter used for the
Rosenbluth separation of cross sections
\eq
\varepsilon=\left[1+\frac{2{\bf q}^2}{Q^2}
\tan^2\frac{\theta_{e^\prime}}{2}\right]^{-1}
=\left[1+\frac{2{{\bf q}^\ast}^2}{Q^2}\tan^2
\frac{\theta^\ast_{e^\prime}}{2}\right]^{-1}
\en 
and $\Gamma_t$ is the ``virtual-photon flux factor''.
In Eq.~(\ref{sig5}) and below the variables defined in the center-of-mass 
reference frame are denoted by $\ast$, while in the lab frame they 
are used without $\ast$, e.g.,
\eq 
& &|{\bf q}|=\sqrt{\left(\frac{W^2\!-\!m_{\scriptscriptstyle N}^2+Q^2}
{2m_{\scriptscriptstyle N}}\right)^2+Q^2},\quad
|{\bf q}^\ast|=\frac{m_{\scriptscriptstyle N}}{W}|{\bf q}|\,, 
\nonumber\\
& &|{\bf q}_r^\ast|=|{\bf q}^\ast|_{Q^2\!=\!0}=
\frac{W^2\!-\!m_{\scriptscriptstyle N}^2}{2W}.
\label{star}
\en 

The final expressions for the longitudinal and transverse cross sections
in the approximation of the lowest order $t-$, $s-$ and $u-$channel diagrams 
(\ref{jt}) read
\eq 
& &\frac{d\sigma_L}{dt} = {\cal N}_\sigma
\frac{1}{4\pi}\overline{\bigg\vert T(\lambda\!=\!0)\biggr\vert^2}\,, 
\nonumber\\ 
& &\frac{d\sigma_T}{dt } = {\cal N}_\sigma
\frac{1}{2}\sum_{\lambda\!=\pm 1}\frac{1}{4\pi} 
\overline{\biggl\vert T(\lambda)\bigg\vert^2}\,,\nonumber\\
& &\frac{d\sigma_{TT}}{dt}={\cal N}_\sigma
\biggl\{-\,\frac{1}{2}\sum_{\lambda\!=\pm 1}\frac{1}{4\pi}
\overline{T(\lambda)T(\!-\!\lambda)^\ast}
\biggr\}\,, \label{fulllt}\\ 
& &\frac{d\sigma_{LT}}{dt} = \nonumber\\ 
& &{\cal N}_\sigma\biggl\{-\,
\frac{1}{2}\sum_{\lambda\!=\pm 1}\lambda
\biggl(\frac{\overline{T(0)T(\lambda)^\ast}
+\overline{T(\lambda)T(0)^\ast}}{4\pi\sqrt{2}}\biggr)
\biggr\}\,, \nonumber 
\en 
where $T(\lambda)$ is the pion electroproduction amplitude, 
which describe the $\pi$-, 
$\rho$-, nucleon-poles- and contact-term contributions to the hadron current
\eq 
& &\langle p^\prime,s^\prime| \epsilon^{(\lambda)}_\mu J^\mu |p,s\rangle\equiv
T(\lambda,s,s^\prime)=T_\pi(\lambda,s,s^\prime) \label{jt}\\
&+&T_\rho^{\mbox{w(v)}}(\lambda,s,s^\prime)
+T_{{\scriptscriptstyle N}(s\!+\!u)}(\lambda,s,s^\prime)
+T_{\scriptscriptstyle CPV}(\lambda,s,s^\prime)\,. 
\nonumber
\en 
The common kinematical factor ${\cal N}_\sigma/4\pi$ is defined by 
the standard expression
\eq
{\cal N}_\sigma=\frac{1}{\sqrt{(W^2-m_{\scriptscriptstyle N}^2+Q^2)^2+
4m_{\scriptscriptstyle N}^2Q^2}}\,\frac{1}{W^2-m_{\scriptscriptstyle N}^2}.
\label{ns}
\en 
The sum over spin projections 
\eq 
\overline{\left\vert T(\lambda)\right\vert^2} =
\frac{1}{2}\sum_{s,s^\prime}\left\vert T(\lambda,s,s^\prime)\right\vert^2 \, 
\label{tmean}
\en 
in Eq.~(\ref{fulllt}) can be calculated by the standard trace technique, 
and the calculation results in an universal formula for all the cross sections 
listed in Eq.~(\ref{fulllt}). In Appendices 
we list the full analytical results for both longitudinal $\sigma_L$ 
and transverse $\sigma_T$ parts given for all the diagonal and 
interference terms.

For the results we have tested two approximations to the $s(u)$-channel 
amplitudes:
\begin{itemize} 
\item[(i)] the naive or ``exact'' representation (\ref{jt}) which makes use 
of the proper Feynman amplitudes (\ref{nsu}) and (\ref{pvcont}) with an 
intermediate virtual nucleon;
\item[(ii)] the effective representation (\ref{simple}) [with two variants,
``soft'' $(a)$ and ``hard'' $(b)$, discussed in Sect.~\ref{sec:formf} below
Eq.~(\ref{lamr})]
taking into account intermediate hadron states through quark diagrams giving 
the main contribution to the ``$s$-channel part'' of the cross section at 
large $Q^2$.
\end{itemize}
Here we furthermore use both the tensor ($W$) and vector ($V$) representation 
of the $\rho$-exchange amplitude.

Our calculation shows that approximation (i) is extremely unrealistic. 
The interference terms between the $t$-pole amplitudes and the ``exact'' 
$T_{{\scriptscriptstyle N}(s\!+\!u)}+T_{\scriptscriptstyle CPV}$ amplitudes 
are too large. The resulting longitudinal cross section is in rather
poor agreement with the observed data on $\sigma_L$. Only the diagonal terms 
without the $s(u)$-pole contributions give qualitative agreement with
the $\sigma_L$ data.

The approximation (ii) is more realistic. Results of this approximation are
displayed in Figs.~\ref{fig5} and \ref{fig6}. By varying the free 
parameter $Z$ one can obtain a good description of both  
cross sections $\sigma_L$ and $\sigma_T$
for the tensor variant of the $\rho$-exchange amplitude 
($W$). For the vector variant ($V$) only $\sigma_L$ can be described 
in agreement with experimental data. However, the ``soft'' variant $(a)$ 
with $\Lambda_{\rm eff}\approx\Lambda_\pi$ used in the form factors 
$F_{\rho\pi\gamma}(t,Q^2)$ and $F_{\rm eff}(Q^2)$ 
is less suitable for describing
the existing data in a large interval of $Q^2$ from 0.6 to 2.45 GeV$^2$/c$^2$,
since it fails to describe the slow $Q^2$ dependence 
of $\sigma_T$ in the full interval. The experimental ratio
$\sigma_T(Q^2{=}0.6)/\sigma_T(Q^2{=}2.45)$ is about 6, while the corresponding
ratio of form factors squared 
$F_{\rho\pi\gamma}^2(Q^2{=}0.6)/F_{\rho\pi\gamma}^2(Q^2{=}2.45)$ 
multiplied by the kinematical factor ${\cal N}_\sigma$ (\ref{ns}) is several
times larger.

Only the ``hard'' variant $(b)$ with $\Lambda_{\rm eff}\gtrsim 2\Lambda_\pi$ is
suitable in describing the data. Taking the value $\Lambda_{\rm eff}=$ 
1.2 GeV/c, which is close to
the conventional values $\Lambda_\rho\approx$ 1 - 1.2 used in literature
for the $\rho\pi\gamma$ form factor (see e.g.~\cite{Horn:2006tm}), and taking 
the standard ``soft'' value $\Lambda_{\rm str}^2=$ 0.7 GeV$^2$/c$^2$ for the 
common strong form factor (\ref{lams})
we obtain a satisfactory description of both $\sigma_L$ and $\sigma_T$ 
cross sections in a wide interval of $Q^2$.

The values of the $\rho NN$ coupling constants were fixed to 
$G_{\rho{\scriptscriptstyle NN}}(0)=$ 4 and 
$F_{\rho{\scriptscriptstyle NN}}(0)=$ 26 close to the recommended in the ChPT 
approach~\cite{Kubis:2000zd} 
of $G_{\rho{\scriptscriptstyle NN}}(0)=$ 4 and 
$F_{\rho{\scriptscriptstyle NN}}(0)/G_{\rho{\scriptscriptstyle NN}}(0)=$ 6.1
(since the value of $\sigma_T$ directly depends on 
$F_{\rho{\scriptscriptstyle NN}}$ we slightly corrected the conventional 
value of $F_{\rho{\scriptscriptstyle NN}}(0)$ to obtain the best description 
of the recent data~\cite{Horn:2006tm} on $\sigma_T$ for 
$Q^2=$ 1.6 GeV$^2$/c$^2$).
In our calculation we only have one free parameter $Z$ introduced in 
Eq.~(\ref{simple}) as a phenomenological constant, which formally corresponds
to a renormalization of the Born contact term 
$T_{\scriptscriptstyle CPV}$ in the full amplitude (\ref{simple}); 
in essence the term $ZF_{\rm eff}(Q^2)F_{\rm str}(t)
T_{\scriptscriptstyle CPV}$ amounts to a phenomenological description of
the $s$-channel contributions, which otherwise cannot be calculated from 
first principles. Based on general considerations 
(see Sect.~\ref{sec:formf}) we can only expect that such contributions are 
suppressed at large $Q^2$, and thus a small value of the phenomenological
constant $Z$ would be expected.  

By varying $Z$ one can improve the
description of only one of the  cross section component 
(on account of the another), $\sigma_L$ or $\sigma_T$. 
To compare our results to the VGL model 
predictions~\cite{Vanderhaeghen:1997ts}, which are only realistic for 
$\sigma_L$, we fit the value of $Z$ to the 
$\sigma_L$ data in the full interval 0.6 $< Q^2<$ 2.45 GeV$^2$/c$^2$. 
With a value of $Z=$ 0.11 we obtain a description
of $\sigma_L$ which practically coincides (see Figs.~\ref{fig5} and 
\ref{fig6}) with the results of the recent Regge-model motivated description 
of the data~\cite{Horn:2006tm,Tadevosyan:2007yd}. 

The small value of $Z$ correlates well with the quark mechanism proposed
in Section~\ref{sec:formf}B for description of $s$-channel contributions
[Figs.~\ref{fig4}(b) and \ref{fig4}(c)] to the cross section. In accordance
with this picture, only a part of all the possible quark diagrams
[Fig.~\ref{fig4}(b)] survives at large $Q^2$, while the most part of
diagrams [of the Fig.~\ref{fig4}(c) type] is suppressed because of a high
degree $n$ of $\sim Q^{-n}$ behavior (it can be assumed that the value of
$Z$ corresponds to the weight $\sim$1/3 of the surviving part of quark
diagrams). On the other hand, it would be very difficult to describe this
situation at large $Q^2$ starting from the interference of many 
baryon-resonance diagrams depicted in Fig.~\ref{fig2}(b). Therefore, the
smallness of $Z$ can be considered as an argument in favor of the 
quark-model motivated representation of the transition amplitude in
Eq.~(\ref{simple}) and against the naive hadron representation in
Eq.~(\ref{jt}).

Following Refs.~\cite{Horn:2006tm,Tadevosyan:2007yd}, 
we use different values for the cutoff parameters $\Lambda_\pi$ in the $F_\pi$ 
form factor for sets of data centered at different values of $Q^2$. 
In Table~1 we compare the values of $F_\pi$ (and correspondingly 
$\Lambda_\pi$) obtained by such a method to the values $F_\pi$ obtained in 
Refs.~\cite{Horn:2006tm,Tadevosyan:2007yd} on the basis of the VGL model. 
The coupling constants and cutoff parameters used in the calculation 
are listed in Table~2. 

In the case of the vector variant $V$, which (as is also the case for the
VGL model) describes only $\sigma_L$ and fails for $\sigma_T$,
the extracted  values of $F_\pi$ are close to the results of 
Refs.~\cite{Horn:2006tm,Tadevosyan:2007yd}. For the tensor variant $W$
our values for $F_\pi$ differ from the ones of 
Ref.~\cite{Horn:2006tm,Tadevosyan:2007yd},
but in this case we obtain a satisfactory description of $\sigma_T$
as a byproduct of our approach.

It is evident that the proper form factor $F_\pi$ cannot fluctuate sharply 
in magnitude from one $Q^2$ to another and the same is true for $\Lambda_\pi$
used for its parametrization. However, after smoothing out the 
fluctuations one can see that there are two $Q^2$ regions, $Q^2\lesssim$
1 GeV$^2$/c$^2$ and $Q^2\gtrsim$ 1 GeV$^2$/c$^2$, with different mean values
for $\Lambda_\pi$. For our $W$ variant we take 
$\Lambda_\pi^2=$ 0.5 GeV$^2$/c$^2$
in the region of smaller $Q^2\le$ 1  GeV$^2$/c$^2$ and 
$\Lambda_\pi^2=$ 0.6 GeV$^2$/c$^2$ in the $Q^2$
region of the recent JLab experiment (in this case in the full interval
the mean value of $\Lambda_\pi^2$ is about 0.54 GeV$^2$/c$^2$) 
and recalculate the cross section
$\sigma_L$ (in practice $\sigma_T$ does not depend on small $F_\pi$ 
variations).
The dotted lines in Figs.~\ref{fig5} and \ref{fig6} show how $\sigma_L(W)$
behaves for  these averaged  values of $F_\pi$.
For comparison, the predicted values of $\Lambda_\pi^2$ for a variety of 
theoretical approaches are shown in Table~3.
 
\section{Summary and conclusions}

We analyzed pion electroproduction $p(e,e^\prime\pi^+)n$, 
which was intensively studied at JLab in the quasi-elastic 
regime~\cite{Volmer:2000ek,Huber:2003kg,Horn:2006tm,Tadevosyan:2007yd} 
for the original purpose of directly measuring the pion charge form factor 
$F_\pi(Q^2)$. 
Our framework is based on an effective Lagrangian approach 
involving nucleon, pion, $\rho$ meson and photon degrees of freedom. 
In the description of the $\rho$ meson we test two possibilities: 
the so-called vector ($V$) and tensor ($W$) variants. 
For the standard vector variant $V$
the transverse part $\sigma_T$ of the cross section is considerably
underestimated, whether it is the Regge model motivated
approach VGL~\cite{Vanderhaeghen:1997ts} or the traditional Born 
one~\cite{Gutbrod:1973qr} modified by strong form factors and contact 
terms~\cite{Neudatchin:2004pu,Obukhovsky:2005sa}. Here we have shown that the
problem of underestimating $\sigma_T$ might be solved by taking into account
specific contact terms in the $\rho$ meson propagator that can be most
naturally obtained in terms of the tensor variant $W$ of the $\rho$ meson
description~\cite{Ecker:1988te,Kubis:2000zd}. 

Based on these findings the  
uniqueness of the $F_\pi$ data extracted from the $\sigma_L$ cross section
appears doubtful
without taking into account the associated data on $\sigma_T$. 
The modified Born approach presented here is successful in the description 
of both the $\sigma_L$ and $\sigma_T$ cross sections, thus present a new 
possibility for the discussion of this problem.

First it should be noted that our results obtained for the standard 
variant $V$ (the dashed lines in Figs.~\ref{fig5} and \ref{fig6} 
for $d\sigma_L/dt$) obviously contradict the claim that the Born 
(i.e. the Feynman tree diagrams) 
approach is completely unsuited for the description of the forward pion 
electroproduction process. In our opinion a refined version of this 
statement would be more adequate: the Born approach is only unsuited for 
the photoproduction ($Q^2=$ 0) and for the low $Q^2$ region of 
electroproduction, but at intermediate and high $Q^2$ this approach becomes 
suitable. This statement is substantiated by a series of previous works~%
\cite{Neudatchin:2004pu,Obukhovsky:2005sa,Yudin:1998bz,Neudatchin:2001ex} 
and in particular, by the present detailed evaluation. 
It seems likely that with growing $Q^2$ the full sum
(not the isolated terms) of $s(u)$-channel contributions is decreasing most and
while the $t$-pole contributions remain. However, any description of such $Q^2$
behavior in terms of many baryon poles would be rather complicated 
and, as a result, very doubtful. 
We can only use physical arguments based on the comparison of
$d\sigma_L/dt$ data to the respective $t$-pole contributions depicted
in Figs.~\ref{fig5} and \ref{fig6} for variant $V$ (the dashed line).
One can see that starting at $Q^2=$ 0.75 GeV$^2$/c$^2$ the $t$ behavior
of the measured $d\sigma_L/dt$ is in a good agreement with predictions
given by the $t$-pole contributions (but  the agreement breaks down
for $Q^2=$ 0.6 GeV$^2$/c$^2$ and for smaller $Q^2$ as our evaluation shows). 

Second, the present successful description of $\sigma_T$ on the basis of the
tensor variant $W$ raises another issue that finally remains to be resolved.
Namely, for the variant $V$, which is very similar in results to the VGL
model predictions, there is no $\pi$-$\rho$ interference contribution
to $\sigma_L$ since the corresponding spin average term vanishes 
(see Appendices). 
But in the more realistic variant $W$ this term does not vanish and cannot be 
neglected. In other words, in a realistic variant of the description of both
$\sigma_L$ and $\sigma_T$ the $\rho$-exchange term influences not only 
$\sigma_T$ but $\sigma_L$ as well. Hence, the procedure of extracting 
$F_\pi$ values from the $\sigma_L$ data alone can, in principle, not be 
independent from 
a parallel description of the $\sigma_T$ data. We evaluated such a possible 
indirect influence of the $\sigma_T$ data on the final $F_\pi$ values extracted
from the recent JLab data (see last two rows in Table~1). In our
simplified model for the $W$ variant, including a phenomenological contact
term (proportional to the free parameter $Z$), we obtained values for 
$F_\pi$ which differ from previous ones (extracted without taking into 
account the $\sigma_T$ data) by about 10 - 15\% 
at $Q^2\gtrsim$ 1.6 GeV$^2$/c$^2$. This deviation is traced to the 
$\rho$ - $\pi$ interference, which cannot be neglected. 
It counts rather in favor of the value 
$\Lambda_\pi^2\approx$ 0.6 GeV$^2$/c$^2$ obtained in our approach than 
in favor of the value $\Lambda_\pi^2\approx$ 0.5 GeV$^2$/c$^2$ obtained
on the basis of the VGL model. However, now we cannot obtain  trustworthy 
values for the uncertainties $\pm \Delta\Lambda_\pi^2$ because of the 
considerable model dependence of the $s$-channel contributions to $\sigma_L$. 

In future we intent to extend our formalism to the study of kaon 
electroproduction in connection with the recent JLAB experiment 
E93-018~\cite{Mohring:2002tr}. 
 
%%%%%%%%%%%%%%%%%%%%%%%%%%%%%%%%%%%%%%%%%%%%%%%%%%%%%%%%%%%%%%%%%%%%%%%%%%%%

\begin{acknowledgments}

This work was supported by the DFG under contracts FA67/31-1, 
436 RUS 113/790 and GRK683. 
This research is also part of the EU Integrated
Infrastructure Initiative Hadronphysics project under contract
number RII3-CT-2004-506078 and President grant of Russia
"Scientific Schools"  No. 5103.2006.2. 
The authors thank the Fpi2 Collaboration 
[Experiment E01-004 at JLab] for the interest to 
our work and informative discussion. 
We thank Vladimir Neudatchin, Dmitri Fedorov, Henk Blok,
Garth Huber and Tanja Horn for fruitful discussions 
and suggestions. 

\end{acknowledgments}

\appendix

%%%%%%%%%%%%%%%%%%%%%%%%%%%%%%%%%%%%%%%%%%%%%%%%%%%%%%%%%%%%%%%%%%%%%%%%%
%%%%%%%%%%%%%%%%%%%%%%%%%%%%%%%%%%%%%%%%%%%%%%%%%%%%%%%%%%%%%%%%%%%%%%

\section{ Longitudinal (L) and transverse (T) cross sections}

%%%%%%%%%%%%%%%%%%%%%%%%%%%%%%%%%%%%%%%%%%%%%%%%%%%%%%%%%%%%%%%%%%%%%%%

We consider the cross sections $d\sigma_{L,\,T}/dt$ integrated over 
the azimuthal angle of the emitted pion $\phi_{\pi^\prime}^\ast$
\eq 
\frac{d\sigma_L}{dt} &=&
\int_0^{2\pi}\frac{d\sigma_L}{dtd\phi_{\pi^\prime}^\ast}
d\phi_{\pi^\prime}^\ast \nonumber\\
&=&\frac{2\pi}{2|{\bf q}^\ast||{\bf q}_r^\ast|}\,\frac{1}{(8\pi W)^2}
\frac{Q^2}{{\bf q}^2}\,\overline{J_0J_0^\dagger},
\label{sigl0}
\en 
\eq
\frac{d\sigma_T}{dt} &=&
\int_0^{2\pi}\frac{d\sigma_T}{dtd\phi_{\pi^\prime}^\ast} \, 
d\phi_{\pi^\prime}^\ast \nonumber\\ 
&=&\frac{2\pi}{2|{\bf q}^\ast||{\bf q}_r^\ast|}\,\frac{1}{(8\pi W)^2}\,
\overline{\left(\frac{J_xJ_x^\dagger\!+\!J_yJ_y^\dagger}{2}\right)} \,. 
\label{siglt}
\en 
For the cross sections $d\sigma_{TT}/dtd\phi_{\pi^\prime}^\ast$ and
$d\sigma_{LT}/dtd\phi_{\pi^\prime}^\ast$, which are proportional to 
$\cos 2\phi_{\pi^\prime}^\ast$ and $\cos \phi_{\pi^\prime}^\ast$ respectively, 
the integral vanishes, and thus the corresponding $d\sigma_{TT,LT}/dt$
values are defined through $d\sigma_{TT,LT}/dtd\phi_{\pi^\prime}^\ast$
at a fixed angle $\phi_{\pi^\prime}^\ast=$ 0 times 2$\pi$ in analogy to 
Eq.~(\ref{siglt}): 
\begin{multline} 
\frac{d\sigma_{TT}}{dt}=
\left.2\pi\frac{d\sigma_{TT}}{dtd\phi_{\pi^\prime}^\ast}
\right\vert_{\phi_{\pi^\prime}^ast\!=0}\\
=\frac{2\pi}{2|{\bf q}^\ast||{\bf q}_r^\ast|}\,\frac{1}{(8\pi W)^2}\,
\overline{\left(\frac{J_xJ_x^\dagger\!-\!J_yJ_y^\dagger}{2}
\right)}_{\phi_{\pi^\prime}^\ast\!=0},
\end{multline}
\begin{multline}
\frac{d\sigma_{LT}}{dt}=
\left.2\pi\frac{d\sigma_{LT}}{dtd\phi_{\pi^\prime}^\ast}
\right\vert_{\phi_{\pi^\prime}^\ast\!=0}\\
=\frac{2\pi}{2|{\bf q}^\ast||{\bf q}_r^\ast|}\,\frac{1}{(8\pi W)^2}\,
\frac{Q}{|{\bf q}|}\,
\overline{\left(-\,\frac{J_0J_x^\dagger\!+\!J_xJ_0^\dagger}
{2}\right)}_{\phi_{\pi^\prime}^\ast\!=0} \,, 
\label{sigttl}
\end{multline} 
The hadron current tensor $J_\mu J_\nu$ averaged and summed over nucleon spin 
projections $s,\,s^\prime$ 
\eq 
\overline{J_\mu J_\nu^\dagger}=
\frac{1}{2}\sum_{s,s^\prime}\langle p^\prime,s^\prime|J_\mu|p,s\rangle\,
[\langle p^\prime,s^\prime|J_\nu|p,s\rangle]^\ast
\en  
can be calculated on the basis of the Feynman matrix elements (\ref{pi}), 
(\ref{rho}), (\ref{nsu}) and (\ref{pvcont}) [see, e.g., Eq.~(\ref{jt})]
which describe the $\pi$-, 
$\rho$-, nucleon-poles- and contact-term contributions to the hadron current

It is important that the Decart (non-invariant) components of the hadron 
tensor, $J_xJ_x^\dagger$, $J_0J_x^\dagger$, etc.,
are presented in Eqs.~(\ref{sigl0}) -- (\ref{sigttl}) in a special 
coordinate frame with the axes $z$ and $y$ directed along the momenta 
${\bf q}$ and ${\bf q}\times{\bf k}^\prime$, respectively. 
Any boost along the $z$-axis does not affect the $x$- and 
$y$-components of the current and, as a result, the transverse components of 
the hadron tensor become invariant with respect to a change of the
reference frame (e.g., from the c.m. system to the lab. system) with
\eq 
\frac{J_xJ_x^\dagger\pm J_yJ_y^\dagger}{2}=
\frac{(J_xJ_x^\dagger)^\ast \pm (J_yJ_y^\dagger)^\ast}{2} \,. 
\en 
Moreover, in the given coordinate frame $\{xyz\}$ there are further 
invariants:
\eq 
& &
\hspace*{-.5cm}
\frac{Q^2}{{\bf q}^2}\,J_0J_0^\dagger=
\frac{Q^2}{{{\bf q}^\ast}^2}\, (J_0J_0^\dagger)^\ast \,,\nonumber\\ 
& &\hspace*{-.5cm}
\frac{Q}{|{\bf q}|}\,[J_0J_x^\dagger+J_xJ_0^\dagger]=
\frac{Q}{|{\bf q}^\ast|}\,[ (J_0J_x^\dagger)^\ast +
(J_xJ_0^\dagger)^\ast] \,. 
\en 
Using the equation
\eq 
\frac{Q}{|{\bf q}|}J_0=J^\mu\epsilon_\mu^{(\lambda\!=0)}(q)\,,
\en
which is valid for a conserved current ($q_\mu J^\mu=$ 0)
one can write the following invariant expressions for all the components
of the hadron tensor of interest:
\eq 
& &\frac{Q^2}{{\bf q}^2}J_0J_0^\dagger=
J^\mu\epsilon_\mu^{(\lambda\!=0)}\,
[J^\nu\epsilon_\nu^{(\lambda\!=0)}]^\dagger\,, \nonumber\\
& &\frac{J_xJ_x^\dagger\pm J_yJ_y^\dagger}{2}
=\pm\frac{1}{2}\sum_{\lambda\!=\pm 1}
J^\mu\epsilon_\mu^{(\lambda)}\,
[J^\nu\epsilon_\nu^{(\pm\lambda)}]^\dagger\,,\\
& &\frac{Q}{|{\bf q}|}
\biggl(\frac{J_0J_x^\dagger\!+\!J_xJ_0^\dagger}{2}\biggr)=\nonumber\\
&-&\frac{1}{2}\sum_{\lambda\!=\pm 1}\frac{\lambda}{\sqrt{2}}
\biggl\{J^\mu\epsilon_\mu^{(0)}\,
[J^\nu\epsilon_\nu^{(\lambda)}]^\dagger
+J^\mu\epsilon_\mu^{(\lambda)}\,
[J^\nu\epsilon_\nu^{(0)}]^\dagger\biggr\} \,. \nonumber
\label{jj}
\en 
Eqs.~(\ref{fulllt}) - (\ref{ns}) of Section \ref{sec:results} give the final 
expressions for the 
longitudinal and transverse cross sections in the approximation of 
the lowest order $t-$, $s-$ and $u-$channel diagrams (\ref{jt}).  
The sum over spin projections (\ref{tmean}) calculated by the standard 
trace technique results in an universal formula for all the cross sections 
listed in Eq.~(\ref{fulllt}). 
Each cross section is expressed through individual dynamical 
factors $\Phi^{\Gamma,\Gamma^\prime}_{pp}(t,Q,W)$, 
$\Phi^{\Gamma,\Gamma^\prime}_{pk}(t,Q,W),\dots$ etc. 
dependent only on the invariants $t$, $Q^2$, $s(=W^2)$
and the polarization factors
$(p\cdot\epsilon^{(\lambda)})$, $(k\cdot\epsilon^{(\lambda)})$, 
$(\epsilon^{(\lambda)}\cdot{\epsilon^{(\lambda)}}^\ast)$ common to all the
cross sections. This can be illustrated for the example of $d\sigma_T/dt$. 
The transverse cross section is decomposed into a sum of partial ones
\eq
\hspace*{-.5cm}
\frac{d\sigma_T}{dt}=\sum_{\scriptscriptstyle\Gamma\Gamma^\prime}
\frac{d\sigma_T^{\scriptscriptstyle\Gamma,\Gamma^\prime}}{dt}\,, \  
\frac{d\sigma_T^{\scriptscriptstyle\Gamma,\Gamma^\prime}}{dt}=
\frac{1}{2}\sum_{\lambda\!=\!\pm 1}\frac{{\cal N}_\sigma}{4\pi}
\overline{T_{\scriptscriptstyle\Gamma}(\lambda)
T_{\scriptscriptstyle\Gamma^\prime}(\lambda)^\ast}
\label{mm1}
\en 
where $\Gamma,\Gamma^\prime=\pi,\,\rho$, $N(s+u)$, $CPV$ refers to the
respective components $T_{\scriptscriptstyle\Gamma}$ and 
$T^\ast_{{\scriptscriptstyle\Gamma}^\prime}$ 
of the full amplitude (\ref{jt}) taken for calculation of the
given partial cross section 
$d\sigma_T^{\scriptscriptstyle\Gamma,\Gamma^\prime}/dt$.
For $\Gamma=\Gamma^\prime$ one obtains the diagonal contribution of the
given mechanism ``$\Gamma$'' to the cross section, while in the
case of $\Gamma\ne \Gamma^\prime$ the partial cross section 
$d\sigma_T^{\Gamma,\Gamma^\prime}/dt$
corresponds to the interference of the amplitudes $T_\Gamma$ and 
$T_{\Gamma^\prime}$. The representation of the final result is of the form
\begin{multline} 
\frac{d\sigma_{\scriptscriptstyle T}}{dt}=
\frac{1}{2}\sum_{\lambda\!=\!\pm 1}\sum_{\Gamma\Gamma^\prime}
\biggl\{(p\cdot\epsilon^{(\lambda)})
(p\cdot\epsilon^{(\lambda)\ast})\Phi^{\Gamma,\Gamma^\prime}_{pp}\\
+\biggl[(p\cdot\epsilon^{(\lambda)})(k\cdot\epsilon^{(\lambda)\ast})
+(p\cdot\epsilon^{(\lambda)\ast})(k\cdot\epsilon^{(\lambda)})\biggr]
\Phi^{\Gamma,\Gamma^\prime}_{pk}\\
+(k\cdot\epsilon^{(\lambda)})
(k\cdot\epsilon^{(\lambda)\ast})\Phi^{\Gamma,\Gamma^\prime}_{kk}
+(\epsilon^{(\lambda)}\cdot\epsilon^{(\lambda)\ast})
\Phi^{\Gamma,\Gamma^\prime}_{ee}
\biggr\}\,,\label{esigt}
\end{multline}
\begin{multline}
\frac{d\sigma_{\scriptscriptstyle L}}{dt}=\sum_{\Gamma\Gamma^\prime}
\biggl\{(p\cdot\epsilon^{(0)})
(p\cdot\epsilon^{(0)\ast})\Phi^{\Gamma,\Gamma^\prime}_{pp}\\
+\biggl[(p\cdot\epsilon^{(0)})(k\cdot\epsilon^{(0)\ast})
+(p\cdot\epsilon^{(0)\ast})(k\cdot\epsilon^{(0)})\biggr]
\Phi^{\Gamma,\Gamma^\prime}_{pk}\\
+(k\cdot\epsilon^{(0)})
(k\cdot\epsilon^{(0)\ast})\Phi^{\Gamma,\Gamma^\prime}_{kk}
+(\epsilon^{(\lambda)}\cdot\epsilon^{(0)\ast})
\Phi^{\Gamma,\Gamma^\prime}_{ee}
\biggr\}\,, \label{esigl}
\end{multline}
containing separately the
polarization components $(\kappa\cdot\epsilon^{(\lambda)}(q)) 
(\kappa^\prime\cdot\epsilon^{(\lambda)\ast}(q))$ 
and the dynamical ones $\Phi^{\Gamma,\Gamma^\prime}_{\kappa\kappa^\prime}$.
In Appendices \ref{sec:appenda}, \ref{sec:appendb} and \ref{sec:appendc} 
we list the full analytical results for both $\sigma_L$ and $\sigma_T$
parts given for all the diagonal and interference terms.

%%%%%%%%%%%%%%%%%%%%%%%%%%%%%%%%%%%%%%%%%%%%%%%%%%%%%%%%%%%%%%%%%%%%%%%%%%

\begin{widetext}

\section{Polarization factors}
\label{sec:appenda}

To further symplify formulas some new dimensionless, invariant variables 
are specified,
\eq 
\eta=\frac{-t}{4m_{\scriptscriptstyle N}^2}, \quad
\xi_s\equiv\frac{p\cdot q}{m_{\scriptscriptstyle N}Q}=
\frac{Q^2+W^2-m_{\scriptscriptstyle N}^2}{2m_{\scriptscriptstyle N}Q},\quad
\xi_t\equiv\frac{k\cdot q}{m_{\scriptscriptstyle N}Q}
=\frac{Q^2-t+M_\pi^2}{2m_{\scriptscriptstyle N}Q}.
\label{par}
\en 

\bigskip

{\it Longitudinal factors :}
\eq 
& &(p\cdot\epsilon^{(\lambda\!=\!0)})(p\cdot
\epsilon^{(\lambda\!=\!0)\,\ast})_{\rm lab}
=m_{\scriptscriptstyle N}^2(1+\xi_s^2),\quad
(k\cdot\epsilon^{(\lambda\!=\!0)})(k\cdot
\epsilon^{(\lambda\!=\!0)\,\ast})_{\rm lab}
=\frac{m_{\scriptscriptstyle N}^2}{1+\xi_s^2}(-2\eta+\xi_s\xi_t)^2, 
\nonumber\\
& &\left[
(p\cdot\epsilon^{(\lambda\!=\!0)})(k\cdot\epsilon^{(\lambda\!=\!0)\,\ast})
+(p\cdot\epsilon^{(\lambda\!=\!0)\,\ast})(k\cdot\epsilon^{(\lambda\!=\!0)})
\right]_{\rm lab}
=2m_{\scriptscriptstyle N}^2(-2\eta+\xi_s\xi_t)\,, \nonumber\\ 
& &(\epsilon^{(\lambda\!=\!0)}\cdot\epsilon^{(\lambda\!=\!0)})=1 \,. 
\label{le}
\en

\bigskip

{\it Transverse factors:}
\eq 
& &\frac{1}{2}\sum_{\lambda=\pm 1}
(p\cdot\epsilon^{(\lambda)})(p\cdot\epsilon^{(\lambda)\,\ast})_{\rm lab}
=0\,,\quad \frac{1}{2}\sum_{\lambda=\pm 1}
\left[(p\cdot\epsilon^{(\lambda)})(k\cdot\epsilon^{(\lambda)\,\ast})
+(p\cdot\epsilon^{(\lambda)\,\ast})
(k\cdot\epsilon^{(\lambda)})\right]_{\rm lab}=0\,,\nonumber\\
& &\frac{1}{2}\sum_{\lambda=\pm 1}
(k\cdot\epsilon^{(\lambda)})(k\cdot\epsilon^{(\lambda)\,\ast})_{\rm lab}
=\frac{-t}{2(1+\xi_s^2)}(1+\eta+\xi_s^2+\xi_t^2-\xi_s\xi_t)
=\frac{1}{2}{\bf k}_{\rm lab}^2 \, \sin^2\theta_k^{\rm lab},\nonumber\\
& &\frac{1}{2}\sum_{\lambda=\pm 1}(\epsilon^{(\lambda)}
\cdot\epsilon^{(\lambda)\,\ast})=-1\,. \label{te}
\en 
Here $\theta_k^{\rm lab}$ is the angle between the vectors ${\bf k}$ and 
${\bf q}$ in the lab. frame:
\eq 
|{\bf k}_{\rm lab}|\cos\theta_k^{\rm lab}
=\frac{m_{\scriptscriptstyle N}}{\sqrt{1+\xi_s^2}}(-2\eta\xi_s-\xi_t)\,, 
\qquad 
|{\bf k}_{\rm lab}|\equiv|{\bf k}|
=\sqrt{-t\left(1-\frac{t}{4m_{\scriptscriptstyle N}^2}\right)}.
\label{cos}
\en

\section{Diagonal contributions to the cross section}
\label{sec:appendb}

\subsection{The diagonal ${\pi}$-meson $t$-pole part}
\eq 
\frac{1}{4\pi}\overline{|T_\pi(\lambda)|^2}&=&8\alpha F_\pi^2(Q^2)
\frac{g_{\pi{\scriptscriptstyle NN}}^2(t)}
{(M_\pi^2-t)^2}
\biggl\{(p\cdot\epsilon^{(\lambda)})(p\cdot\epsilon^{(\lambda)\,\ast})P_{pp}
+(k\cdot\epsilon^{(\lambda)})(k\cdot\epsilon^{(\lambda)\,\ast})P_{kk}
\nonumber\\
&+&\left[(p\cdot\epsilon^{(\lambda)})(k\cdot\epsilon^{(\lambda)\,\ast})
+(p\cdot\epsilon^{(\lambda)\,\ast})(k\cdot\epsilon^{(\lambda)})\right]P_{pk}
+(\epsilon^{(\lambda)}\cdot\epsilon^{(\lambda)\,\ast})P_{ee}\biggr\}\,,
\label{spi}
\en 

{\it Invariant factors} 
$P_{\kappa\kappa^\prime}\,
(\equiv\Phi^{\pi\pi}_{\kappa\kappa^\prime}(t,Q,W)\,)$
are the same for PS and PV couplings:
\eq 
P_{kk}=-t,\qquad P_{pp}=P_{pk}=P_{ee}=0\,.
\label{ifp}
\en 

\subsection{The diagonal ${\rho}$-meson $t$-pole part}
\eq 
\frac{1}{4\pi}
\overline{|T_\rho^{V(W)}(\lambda)|^2}&=& 
\frac{\alpha}{2} \frac{g_{\rho\pi\gamma}^2(t,Q^2)} 
{(M_\rho^2-t)^2} m_{\scriptscriptstyle N}^2 
\biggl\{(p\cdot\epsilon^{(\lambda)})(p\cdot\epsilon^{(\lambda)\,\ast})
R_{pp}^{V(W)}
+(k\cdot\epsilon^{(\lambda)})(k\cdot\epsilon^{(\lambda)\,\ast})
R_{kk}^{V(W)}\nonumber \\
&+&\biggl[(p\cdot\epsilon^{(\lambda)})(k\cdot\epsilon^{(\lambda)\,\ast})
+(p\cdot\epsilon^{(\lambda)\,\ast})(k\cdot\epsilon^{(\lambda)})\biggr]
R_{pk}^{V(W)}
+(\epsilon^{(\lambda)}\cdot\epsilon^{(\lambda)\,\ast})
R_{ee}^{V(W)}\biggr\}\,.
\label{rdlv}
\en 

{\it Invariant factors} $R^{V(W)}_{\kappa\kappa^\prime}\,
(\equiv\Phi^{\rho\rho}_{\kappa\kappa^\prime}(t,Q,W)\,)$:
\eq 
R_{pp}^V&=&4Q^2\biggl[-(\xi_t^2-4\eta)\biggl(G_\rho^2 
+ \eta F_\rho^2\biggr)\biggr]\,, \\
R_{pp}^W&=&R_{pp}^V+4Q^2
\biggl(\frac{M_\rho^2-t}{m_{\scriptscriptstyle N}^2}\biggr)
\biggl(\frac{1}{2}\xi_t^2+\frac{M_\rho^2+t}{4m_{\scriptscriptstyle N}^2}
\biggr)F_\rho^2\,, \\ 
R_{pk}^V&=&4Q^2\biggl[(\xi_s\xi_t-2\eta)
\biggl(G_\rho^2+\eta F_\rho^2\biggr)
\biggr]\,, \\
R_{pk}^W&=&R_{pk}^{\mbox{v}}+4Q^2
\biggl(\frac{M_\rho^2-t}{m_{\scriptscriptstyle N}^2}\biggr)
\biggl[-\frac{1}{2}\biggl(
\xi_s\xi_t+\frac{M_\rho^2+t}{4m_{\scriptscriptstyle N}^2}
\biggr)\biggr]F_\rho^2\,, \\
R_{kk}^V&=&4Q^2\biggl[-\xi_s^2\biggl(G_\rho^2+\eta F_\rho^2\biggr)
+2\eta G_\rho F_\rho-(1-\eta)G_\rho^2\biggr]\,,\\
R_{kk}^W&=&R_{kk}^{\mbox{v}}+4Q^2
\biggl(\frac{M_\rho^2-t}{m_{\scriptscriptstyle N}^2}\biggr)
\biggl[\frac{1}{2}(\xi_s^2F_\rho^2-G_\rho F_\rho)\biggr]\,, \\
R_{ee}^V&=&4Q^2m_{\scriptscriptstyle N}^2\biggl\{
\biggl[\xi_t^2-4\eta\xi_s(\xi_s-\xi_t)\biggr]
\biggl(G_\rho^2+\eta F_\rho^2\biggr)-\eta\xi_t^2(G_\rho+F_\rho)^2
-4\eta(G_\rho-\eta F_\rho)^2\biggr\}\,, \\
R_{ee}^W&=&R_{ee}^V+4Q^2m_{\scriptscriptstyle N}^2
\biggl(\frac{M_\rho^2-t}{m_{\scriptscriptstyle N}^2}\biggr)
\biggl\{-\,\frac{M_\rho^2+t}{4m_{\scriptscriptstyle N}^2}
\biggl[\xi_s(\xi_s-\xi_t)+\eta\biggr]F_\rho^2
+\frac{1}{2}(\xi_t^2-4\eta)G_\rho F_\rho\biggr\}\,.  
\en 
Here for simplicity we define $G_\rho=G_{\rho NN}(t)$ and 
$F_\rho=F_{\rho NN}(t)$.

\subsection{The diagonal nucleon ($s$- and $u$-pole) part}

\eq 
\frac{1}{4\pi}\overline{|T_{{\scriptscriptstyle N}(s+u)}(\lambda)|^2}&=&
2\alpha \frac{g_{\pi{\scriptscriptstyle NN}}^2F_{\rm eff}^2(t,Q^2)}
{(W^2-m_{\scriptscriptstyle N}^2)^2}
\biggl\{(p\cdot\epsilon^{(\lambda)})(p\cdot\epsilon^{(\lambda)\,\ast})U_{pp}
+(k\cdot\epsilon^{(\lambda)})(k\cdot\epsilon^{(\lambda)\,\ast})U_{kk} 
\nonumber\\
&+&\biggl[(p\cdot\epsilon^{(\lambda)})(k\cdot\epsilon^{(\lambda)\,\ast})
+(p\cdot\epsilon^{(\lambda)\,\ast})(k\cdot\epsilon^{(\lambda)})\biggr]U_{pk}
+(\epsilon^{(\lambda)}\cdot\epsilon^{(\lambda)\,\ast})U_{ee}
\biggr\}\,. 
\label{snsu}
\en 

\noindent
{\it Invariant factors} $U_{\kappa\kappa^\prime}\,
(\,\equiv\Phi^{\scriptscriptstyle NN}_{\kappa\kappa^\prime}(t,Q,W)\,)$:

\eq 
U_{pp}&=&4m_{\scriptscriptstyle N}^2\biggl\{
-\frac{M_\pi^2}{m_{\scriptscriptstyle N}^2}
\biggr[(F_{1p}-\widetilde {F}_{1n})^2
+\frac{Q^2}{4m_{\scriptscriptstyle N}^2}(F_{2p}-\widetilde {F}_{2n})^2\biggr]
+\frac{Q^2}{m_{\scriptscriptstyle N}^2}
\biggl(\xi_t-\frac{Q}{m_{\scriptscriptstyle N}}\biggr)^2
F_{2p}\widetilde {F}_{2n}
\biggr\}\,, \label{upp}\\
U_{pk}&=&4m_{\scriptscriptstyle N}^2\biggl\{
\frac{Q}{m_{\scriptscriptstyle N}}
\biggl(\xi_s-\frac{Q}{2m_{\scriptscriptstyle N}}\biggr)
\biggl[(F_{1p}-\widetilde{F}_{1n})^2
+\frac{Q^2}{4m_{\scriptscriptstyle N}^2}(F_{2p}-\widetilde{F}_{2n})^2
-\frac{Q}{m_{\scriptscriptstyle N}}
\biggl(\xi_t-\frac{Q}{m_{\scriptscriptstyle N}}\biggr)
F_{2p}\widetilde{F}_{2n}\biggr] \nonumber\\
&-&\biggl(\frac{Q}{m_{\scriptscriptstyle N}}\xi_t-4\eta\biggr)
\biggl[(F_{1p}-\widetilde {F}_{1n})\widetilde {F}_{1n}
+\frac{Q^2}{4m_{\scriptscriptstyle N}^2}(F_{2p}-\widetilde{F}_{2n})
\widetilde {F}_{2n} \biggr]\biggr\} \,,\label{upk}\\
U_{kk}&=&4m_{\scriptscriptstyle N}^2\biggl\{
\frac{2Q}{m_{\scriptscriptstyle N}}\xi_s
\biggl[(F_{1p}-\widetilde{F}_{1n})\widetilde{F}_{1n}
+\frac{Q^2}{4m_{\scriptscriptstyle N}^2}(F_{2p}-\widetilde{F}_{2n})
\widetilde{F}_{2n}\biggr]
+4\eta\biggl(\widetilde{F}_{1n}^2
+\frac{Q^2}{4m_{\scriptscriptstyle N}^2}\widetilde{F}_{2n}^2\biggr)
\nonumber\\
&+&\frac{Q^2}{m_{\scriptscriptstyle N}^2}
\biggr[\biggl(\xi_s^2-\frac{Q}{m_{\scriptscriptstyle N}}\xi_s+1\biggr)
F_{2p}\widetilde{F}_{2n}+(F_{1p}\widetilde{F}_{2n}+\widetilde{F}_{1n}F_{2p})
\biggr]\biggr\}\,,\label{ukk} \\
U_{ee}&=&4m_{\scriptscriptstyle N}^2Q^2\biggl\{
\biggl[\frac{Q}{2m_{\scriptscriptstyle N}}\biggl(\xi_t
-\frac{Q}{2m_{\scriptscriptstyle N}}\biggr)
-\eta\biggl(\xi_s-\frac{Q}{2m_{\scriptscriptstyle N}}\biggr)^2\biggr]
(F_{2p}-\widetilde{F}_{2n})^2 \nonumber\\
&+&\xi_s(\xi_t-\xi_s)(F_{1p}-\widetilde{F}_{1n})^2
-\biggl[\frac{Q}{2m_{\scriptscriptstyle N}}(2\xi_s-\xi_t)-\xi_s\xi_t\biggr]
(F_{1p}-\widetilde{F}_{1n})(F_{2p}-\widetilde{F}_{2n}) \nonumber\\
&-&2\biggl[\frac{Q}{2m_{\scriptscriptstyle N}}(2\xi_s-\xi_t)-\xi_s\xi_t
+\frac{1}{2}\xi_t^2\biggr]
\biggl[(F_{1p}-\widetilde{F}_{1n})-\eta(F_{2p}-\widetilde{F}_{2n})
\biggr]\widetilde{F}_{2n} \nonumber\\
&-&\,(1+\eta)\left(\xi_t
-\frac{Q}{m_{\scriptscriptstyle N}}\right)^2F_{2p}\widetilde{F}_{2n}
-\eta(F_{1p}-\widetilde{F}_{1n}+F_{2p}-\widetilde{F}_{2n})^2
\biggr\}\,,\label{uee}
\en 
\noindent
where $F_{1p}$, $F_{2p}$, $F_{1n}$ and $F_{2n}$ are the Dirac nucleon form 
factors and we denote for simplicity
\eq 
\widetilde {F}_{in}=\frac{W^2-m_{\scriptscriptstyle N}^2}
{W^2-m_{\scriptscriptstyle N}^2+Q^2+t-M_\pi^2}{F}_{in},\quad 
F_{ip}=F_{ip}(Q^2),\quad F_{in}=F_{in}(Q^2),\quad i=1,\,2.
\label{fn}
\en 

\subsection{The diagonal part of the contact term contribution}
\eq 
\frac{1}{4\pi}\overline{|T_{\scriptscriptstyle CPV}(\lambda)|^2}
&=&\alpha F_\pi^2(Q^2)
\frac{g_{\pi{\scriptscriptstyle NN}}^2(t)}{2m_{\scriptscriptstyle N}^2}
\biggl\{(p\cdot\epsilon^{(\lambda)})(p\cdot\epsilon^{(\lambda)\,\ast})C_{pp}
+(k\cdot\epsilon^{(\lambda)})(k\cdot\epsilon^{(\lambda)\,\ast})C_{kk} 
\nonumber\\
&+&\biggl[(p\cdot\epsilon^{(\lambda)})(k\cdot\epsilon^{(\lambda)\,\ast})
+(p\cdot\epsilon^{(\lambda)\,\ast})(k\cdot\epsilon^{(\lambda)})\biggr]C_{pk}
+(\epsilon^{(\lambda)}\cdot\epsilon^{(\lambda)\,\ast})C_{ee}
\biggr\}
\label{spiga}
\en 

\noindent
{\it Invariant factors} $C_{\kappa\kappa^\prime}\,(\equiv
\Phi^{\pi\gamma{-}\pi\gamma}_{\kappa\kappa^\prime}(t,Q,W)\,)$:
\eq 
C_{pp}=4,\qquad C_{pk}=-2,\qquad C_{kk}=0,\qquad 
C_{ee}=-4m_{\scriptscriptstyle N}^2(1+\eta).
\label{cpiga}
\en

\section{Interference terms}
\label{sec:appendc}

\subsection{Interference terms for the ${\bm\pi}$- and  ${\bm\rho}$-meson poles}
\eq 
& &\frac{1}{4\pi}\overline{\biggl[T_{\pi}(\lambda)
T_{\rho}^{V(W)\ast}(\lambda)
+T_{\rho}^{V(W)}(\lambda)T_{\pi}^\ast(\lambda)\biggr]} 
= \nonumber\\
&=&\alpha m_{\scriptscriptstyle N} F_\pi(Q^2)
\frac{g_{\pi{\scriptscriptstyle NN}}(t)}
{M_\pi^2-t}\frac{g_{\rho\pi\gamma}(t,Q^2)}{M_\rho^2-t}
\biggl\{(p\cdot\epsilon^{(\lambda)})(p\cdot\epsilon^{(\lambda)\,\ast})
T_{pp}^{V(W)} 
+(k\cdot\epsilon^{(\lambda)})(k\cdot\epsilon^{(\lambda)\,\ast}) 
T_{kk}^{V(W)}\nonumber\\
&+&\biggl[(p\cdot\epsilon^{(\lambda)})(k\cdot\epsilon^{(\lambda)\,\ast})
+(p\cdot\epsilon^{(\lambda)\,\ast})(k\cdot\epsilon^{(\lambda)})\biggr]
T_{pk}^{V(W)}
+(\epsilon^{(\lambda)}\cdot\epsilon^{(\lambda)\,\ast})T_{ee}^{V(W)}
\biggr\}\,.
\label{pirho}
\en 

\noindent
{\it Invariant factors} $T^{V(W)}_{\kappa\kappa^\prime}\,
(\equiv\Phi^{\pi\rho}_{\kappa\kappa^\prime}(t,Q,W)\,)$ 
are the same for PS- and PV couplings:
\eq 
& &T_{pk}^{W}=
-\,2Qm_{\scriptscriptstyle N}
\biggl(\frac{M_\rho^2-t}{m_{\scriptscriptstyle N}^2}\biggr)\xi_tF_\rho(t)\,,
\qquad T_{kk}^{W}=4Qm_{\scriptscriptstyle N}
\biggl(\frac{M_\rho^2-t}{m_{\scriptscriptstyle N}^2}\biggr)
\xi_s F_\rho(t)\,,\nonumber\\
& &T_{pp}^{V}=T_{pk}^{V}=T_{kk}^{V}=T_{ee}^{V}=0,
\hspace*{1.9cm} T_{pp}^{W}=T_{ee}^{W}=0\,. 
\label{tpirho}
\en

\subsection{Interference terms for the ${\bm\pi}$-meson ($t$) and 
nucleon ($s+u$) poles}
\eq 
& &\frac{1}{4\pi}\overline{\biggl[T_{\pi}(\lambda)
T_{{\scriptscriptstyle N}(s+u)}^\ast(\lambda)
+T_{{\scriptscriptstyle N}(s+u)}(\lambda)T_{\pi}^\ast(\lambda)\biggr]} = 
\nonumber\\
&=& - 2 \alpha F_\pi(Q^2)  
\frac{g_{\pi{\scriptscriptstyle NN}}(t)}{M_\pi^2-t}
\frac{g_{\pi{\scriptscriptstyle NN}}F_{\rm eff}(t,Q^2)}
{W^2-m_{\scriptscriptstyle N}^2}
\biggl\{(p\cdot\epsilon^{(\lambda)})(p\cdot\epsilon^{(\lambda)\,\ast})
N_{pp} 
+(k\cdot\epsilon^{(\lambda)})(k\cdot\epsilon^{(\lambda)\,\ast})N_{kk} 
\nonumber\\ 
&+&\biggl[(p\cdot\epsilon^{(\lambda)})(k\cdot\epsilon^{(\lambda)\,\ast})
+(p\cdot\epsilon^{(\lambda)\,\ast})(k\cdot\epsilon^{(\lambda)})\biggr]N_{pk}
+(\epsilon^{(\lambda)}\cdot\epsilon^{(\lambda)\,\ast})N_{ee}
\biggr\} 
\label{ipnl}
\en 

\noindent
{\it Invariant factors}
$N_{\kappa\kappa^\prime}\,
(\equiv\Phi^{\pi{\scriptscriptstyle N}}_{\kappa\kappa^\prime}(t,Q,W)\,)$ 
are different for PS and PV couplings

1) PS coupling:
\eq 
& &N_{pp}=N_{ee}=0\,, \qquad\qquad 
N_{pk}=-4m_{\scriptscriptstyle N}^2
\biggl(4\eta-\frac{Q}{m_{\scriptscriptstyle N}}\xi_t\biggr)
\biggl(F_{1p}-\widetilde{F}_{1n}\biggr),\nonumber\\ 
& &N_{kk}=-8m_{\scriptscriptstyle N}^2\biggl[
4\eta\widetilde{F}_{1n}+\frac{Q}{m_{\scriptscriptstyle N}}\xi_s 
(F_{1p}-\widetilde{F}_{1n})
+\frac{Q^2}{2m_{\scriptscriptstyle N}^2}(F_{2p}+\widetilde {F}_{2n})
\biggr] \label{ipnts}
\en 
2) PV coupling:
\eq 
N_{pp}&=&N_{ee} \, = \, 0\,,\\
N_{pk}&=&-4m_{\scriptscriptstyle N}^2\biggl[4\eta(F_{1p}-\widetilde{F}_{1n})
-\frac{Q}{m_{\scriptscriptstyle N}}\xi_t\biggl(
(F_{1p}-\widetilde{F}_{1n})-\frac{Q^2}{4m_{\scriptscriptstyle N}^2}
(F_{2p}-\widetilde{F}_{2n})\biggr)\nonumber\\
&-&\frac{Q^2}{2m_{\scriptscriptstyle N}^2}\xi_t\biggl(
\xi_s(F_{2p}+\widetilde{F}_{2n})-\xi_t\widetilde{F}_{2n}
\biggr)\biggr]\,, \\
N_{kk}&=&-8m_{\scriptscriptstyle N}^2\biggl[
(4\eta-\frac{Q}{m_{\scriptscriptstyle N}}\xi_t)\,\widetilde{F}_{1n}
-\frac{Q}{m_{\scriptscriptstyle N}}\xi_s
\frac{Q^2}{4m_{\scriptscriptstyle N}^2}(F_{2p}-\widetilde{F}_{2n})
\nonumber\\
&+&\frac{Q^2}{2m_{\scriptscriptstyle N}^2}\xi_s\biggl(
\xi_s(F_{2p}+\widetilde{F}_{2n})-\xi_t\widetilde{F}_{2n}\biggr)
+\frac{Q^2}{2m_{\scriptscriptstyle N}^2}
(F_{1p}+\widetilde{F}_{1n}+F_{2p}+\widetilde {F}_{2n})
\biggr] \,. 
\label{ipntv}
\en 

\subsection{Interference term for the ${\bm\pi}$-meson pole and 
the contact ${\bm\gamma}{\bm\pi}{\bf NN}$ vertex}
\eq 
& &\frac{1}{4\pi}\overline{\biggl[T_{\pi}(\lambda)
T_{\scriptscriptstyle CPV}^\ast(\lambda)
+T_{\scriptscriptstyle CPV}(\lambda)T_{\pi}^\ast(\lambda)\biggr]}= 
\nonumber\\
&=&\alpha F_\pi^2(Q^2)\frac{g_{\pi{\scriptscriptstyle NN}}^2(t)}
{2m_{\scriptscriptstyle N}^2 \, M_\pi^2-t}
\biggl\{(p\cdot\epsilon^{(\lambda)})
(p\cdot\epsilon^{(\lambda)\,\ast})P_{pp}
+\biggl[(p\cdot\epsilon^{(\lambda)})(k\cdot\epsilon^{(\lambda)\,\ast})
+(p\cdot\epsilon^{(\lambda)\,\ast})(k\cdot\epsilon^{(\lambda)})\biggr]P_{pk}
\nonumber\\
&+&(k\cdot\epsilon^{(\lambda)})(k\cdot\epsilon^{(\lambda)\,\ast})P_{kk}
+(\epsilon^{(\lambda)}\cdot\epsilon^{(\lambda)\,\ast})P_{ee}
\biggr\}
\label{ppg}
\en 
{\it Invariant factors} $P_{\kappa\kappa^\prime}\,(\equiv
\Phi^{\pi{-}{\scriptscriptstyle CPV}}_{\kappa\kappa^\prime}(t,Q,W)\,)$:
\eq 
P_{pp}=P_{pk}=P_{ee}=0,\qquad P_{kk}=16m_{\scriptscriptstyle N}^2 \,.
\label{ppiga}
\en 

\subsection{Interference terms for the ${\bm\rho}$-meson ($t$) and 
nucleon ($s+u$) poles}
\eq 
& &\frac{1}{4\pi}\overline{\biggl[T_{{\scriptscriptstyle N}(s+u)}(\lambda)
T_{\rho}^{V(W)\ast}(\lambda)
+T_{\rho}^{V(W)}(\lambda)
T_{{\scriptscriptstyle N}(s+u)}^\ast(\lambda)\biggr]} = \nonumber\\
&=& - \alpha m_{\scriptscriptstyle N} 
\frac{g_{\rho\pi\gamma}(t,Q^2)}{M_\rho^2-t}  
\frac{g_{\pi{\scriptscriptstyle NN}}F_{\rm eff}(t,Q^2)}
{W^2-m_{\scriptscriptstyle N}^2}\, 
\biggl\{(p\cdot\epsilon^{(\lambda)})(p\cdot\epsilon^{(\lambda)\,\ast})
H_{pp}^{\{\mbox{v,w}\}}
+(k\cdot\epsilon^{(\lambda)})(k\cdot\epsilon^{(\lambda)\,\ast})
H_{kk}^{V(W)}\nonumber\\ 
&+&\biggl[(p\cdot\epsilon^{(\lambda)})(k\cdot\epsilon^{(\lambda)\,\ast})
+(p\cdot\epsilon^{(\lambda)\,\ast})(k\cdot\epsilon^{(\lambda)})\biggr]
H_{pk}^{V(W)} + (\epsilon^{(\lambda)}\cdot\epsilon^{(\lambda)\,\ast})
H_{ee}^{V(W)}
\biggr\}\,. 
\label{rnv}
\en 
\noindent
{\it Invariant factors}
$H^{V(W)}_{\kappa\kappa^\prime}\,
(\equiv\Phi^{\rho{\scriptscriptstyle N}}_{\kappa\kappa^\prime}(t,Q,W)\,)$

1) PS coupling: 
\eq 
H_{pp}^{V}&=&4Q^2(\xi_t^2-4\eta)\biggl[ 
\biggl(F_{1p}-\widetilde{F}_{1n}\biggr)F_\rho
-\biggl(F_{2p}-\widetilde{F}_{2n}\biggr)G_\rho\biggr]\,,\\[-1.5mm]
H_{pp}^{W}&=&H_{pp}^{V}+4Qm_{\scriptscriptstyle N}
\biggl(\frac{M_\rho^2-t}{m_{\scriptscriptstyle N}^2}\biggr)
\biggl(\frac{Q}{m_{\scriptscriptstyle N}}-\xi_t\biggr)
\biggl(F_{1p}-\widetilde{F}_{1n}\biggr)F_\rho\,, \label{vppls}\\[-1.5mm]
H_{pk}^{V}&=&4Q^2\biggl\{-\xi_s\xi_t\biggl[
\biggl(F_{1p}-\widetilde{F}_{1n}\biggr)F_\rho
-\biggl(F_{2p}-\widetilde{F}_{2n}\biggr)G_\rho\biggr]
-\xi_t\frac{Q}{4m_{\scriptscriptstyle N}}
\biggl(F_{2p}+\widetilde{F}_{2n}\biggr)(G_\rho+F_\rho)\nonumber\\[-1.5mm]
&+&\eta\biggl[\biggl(F_{2p}-\widetilde{F}_{2n}\biggr)(F_\rho-G_\rho)
+2\biggl(F_{1p}-\widetilde{F}_{1n}\biggr)F_\rho
+2\widetilde{F}_{2n}(G_\rho+F_\rho)\biggr]\biggr\}\,,\\[-1.5mm]
H_{pk}^{W}&=&H_{pk}^{V}+4Qm_{\scriptscriptstyle N}
\biggl(\frac{M_\rho^2-t}{m_{\scriptscriptstyle N}^2}\biggr)
\biggl[\biggl(\frac{1}{2}\xi_s-\frac{Q}{2m_{\scriptscriptstyle N}}\biggr)
\biggl(F_{1p}-\widetilde{F}_{1n}\biggr) 
-\frac{1}{2}\xi_t\widetilde{F}_{1n}  
-\frac{Q}{4m_{\scriptscriptstyle N}}\biggl(F_{2p}+\widetilde{F}_{2n}\biggr)
\biggr]F_\rho\,,
\label{vpkls} \\[-1.5mm]
H_{kk}^{V}&=&4Q^2\biggl[
\biggl(F_{1p}-\widetilde{F}_{1n}\biggr)\biggl(\xi_s^2F_\rho-G_\rho\biggr)
-\biggl(1+\xi_s^2\biggr)\biggl(F_{2p}-\widetilde{F}_{2n}\biggr)G_\rho
\nonumber\\[-1.5mm]
&+&\xi_s\frac{Q}{2m_{\scriptscriptstyle N}}
\biggl(F_{2p}+\widetilde{F}_{2n}\biggr)(G_\rho+F_\rho)
-2\eta\widetilde{F}_{2n}(G_\rho+F_\rho)\biggr]\,,\\
H_{kk}^{W}&=&H_{kk}^{\mbox{v}}+4Qm_{\scriptscriptstyle N}
\biggl(\frac{M_\rho^2-t}{m_{\scriptscriptstyle N}^2}\biggr)
\biggl(\xi_s\widetilde{F}_{1n}
+\frac{Q}{2m_{\scriptscriptstyle N}}\widetilde{F}_{2n}\biggr)F_\rho\,, 
\label{vkkls} \\[-1.5mm]
H_{ee}^{V}&=&4Q^2m_{\scriptscriptstyle N}^2
\biggl\{\eta\biggl[\biggl(\xi_t-2\xi_s\biggr)
\biggl(\xi_t-\frac{Q}{m_{\scriptscriptstyle N}}\biggr)
\biggl(F_{2p}+\widetilde{F}_{2n}\biggr)(G_\rho+F_\rho)
-\xi_t^2\biggl(F_{2p}-\widetilde{F}_{2n}\biggr)
(G_\rho+F_\rho)\nonumber\\[-1.5mm]
&+&4\xi_s\biggl(\xi_s-\xi_t\biggr)
\biggl((F_{1p}-\widetilde{F}_{1n})F_\rho-
(F_{2p}-\widetilde{F}_{2n})G_\rho\biggr)
-4(F_{1p}-\widetilde{F}_{1n}+F_{2p}-\widetilde{F}_{2n})
(G_\rho-\eta F_\rho)\biggr]\nonumber\\[-1.5mm]
&+&\xi_t^2(F_{1p}-\widetilde{F}_{1n}+F_{2p}-\widetilde{F}_{2n})G_\rho
\biggr\}\,,\\[-1.5mm]
H_{ee}^{W}&=&H_{ee}^{\mbox{v}}+4Q^2m_{\scriptscriptstyle N}^2
\biggl(\frac{M_\rho^2-t}{m_{\scriptscriptstyle N}^2}\biggr)
\biggl\{\biggl[
-\eta\biggl(F_{1p}-\widetilde{F}_{1n}+F_{2p}-\widetilde{F}_{2n}\biggr)
-\frac{Q}{4m_{\scriptscriptstyle N}}\biggl(2\xi_s-\xi_t\biggr)
\biggl(F_{2p}+\widetilde{F}_{2n}\biggr)\nonumber\\[-1.5mm]
&-&\xi_s\biggl(\xi_s-\xi_t\biggr)\biggl(F_{1p}-\widetilde{F}_{1n}\biggr)
+\frac{1}{2}\xi_s\xi_t\biggl(F_{2p}+\widetilde{F}_{2n}\biggr)
-\frac{1}{2}\xi_t^2\widetilde{F}_{2n}\biggr]F_\rho\biggr\}\,.
\label{veels} 
\en

2) PV coupling: 
\eq 
H_{pp}^{V}&=&4Q^2\biggl(\xi_t^2-4\eta\biggr)
\biggl[(F_{1p}-\widetilde{F}_{1n})F_\rho-(F_{2p}-\widetilde{F}_{2n})
\biggl(G_\rho+\frac{Q^2}{4m_{\scriptscriptstyle N}^2}F_\rho\biggr)
\nonumber\\
&+&\frac{Q}{2m_{\scriptscriptstyle N}}
\biggl(\xi_s(F_{2p}+\widetilde{F}_{2n})-\xi_t\widetilde{F}_{2n}
\biggr)F_\rho\biggr]\,,\\
H_{pp}^{W}&=&H_{pp}^{V}+4Qm_{\scriptscriptstyle N}
\biggl(\frac{M_\rho^2-t}{m_{\scriptscriptstyle N}^2}\biggr)
\biggl\{\biggl[\biggl(\frac{Q}{m_{\scriptscriptstyle N}}-\xi_t\biggr)
(F_{1p}-\widetilde{F}_{1n})-\frac{Q^2}{4m_{\scriptscriptstyle N}^2}
\frac{Q}{m_{\scriptscriptstyle N}}(F_{2p}-\widetilde{F}_{2n})
\biggr]\nonumber\\
&+&\frac{Q^2}{2m_{\scriptscriptstyle N}^2}
\biggl(\xi_s(F_{2p}+\widetilde{F}_{2n})-\xi_t\widetilde{F}_{2n}
\biggr)\biggr\}F_\rho \,. \label{vpplv} 
\en 
\eq
H_{pk}^{V}&=&2Q^2\biggl\{-(2\xi_s\xi_t-4\eta)
\biggl[(F_{1p}-\widetilde{F}_{1n})F_\rho-(F_{2p}-\widetilde{F}_{2n})
\biggl(G_\rho+\frac{Q^2}{4m_{\scriptscriptstyle N}^2}F_\rho\biggr)\biggr]
\nonumber\\
&-&\frac{Q}{2m_{\scriptscriptstyle N}}(2\xi_s\xi_t-4\eta)
\biggl(\xi_s(F_{2p}+\widetilde{F}_{2n})-\xi_t\widetilde{F}_{2n}\biggr)F_\rho
-\frac{Q}{2m_{\scriptscriptstyle N}}\xi_t
(F_{1p}+\widetilde{F}_{1n}+F_{2p}+\widetilde{F}_{2n})(G_\rho+F_\rho)
\nonumber\\
&+&\biggl[\xi_s\xi_t(F_{1p}-\widetilde{F}_{1n})
+\xi_t^2\widetilde{F}_{1n}+2\eta(F_{2p}+\widetilde{F}_{2n})
\biggr](G_\rho+F_\rho)\biggr\},\\
H_{pk}^{W}&=&H_{pk}^{V}+2Qm_{\scriptscriptstyle N}
\biggl(\frac{M_\rho^2-t}{m_{\scriptscriptstyle N}^2}\biggr)
\bigg\{-\biggl[\biggl(\frac{Q}{m_{\scriptscriptstyle N}}-\xi_s\biggr)
(F_{1p}-\widetilde{F}_{1n})-\frac{Q^2}{4m_{\scriptscriptstyle N}^2}
\frac{Q}{m_{\scriptscriptstyle N}}(F_{2p}-\widetilde{F}_{2n})\biggr]
\nonumber\\
&-&\frac{Q^2}{2m_{\scriptscriptstyle N}^2}
\biggl(\xi_s(F_{2p}+\widetilde{F}_{2n})-\xi_t\widetilde{F}_{2n}\biggr)
-\frac{Q}{2m_{\scriptscriptstyle N}}(F_{2p}+\widetilde{F}_{2n})
-\xi_t\widetilde{F}_{1n}\biggr\}F_\rho \,.
\label{wpklv} \\
H_{kk}^{V}&=&4Q^2\biggl\{\frac{Q}{2m_{\scriptscriptstyle N}}\xi_s
\biggl[(F_{1p}+\widetilde{F}_{1n})(G_\rho+F_\rho)
+(1+\xi_s^2)(F_{2p}+\widetilde{F}_{2n})F_\rho\biggr]\nonumber\\
&+&\biggl[\frac{Q}{2m_{\scriptscriptstyle N}}\xi_t\widetilde{F}_{2n}
+\frac{Q^2}{4m_{\scriptscriptstyle N}^2}(F_{2p}-\widetilde{F}_{2n})\biggr]
(G_\rho-\xi_s^2F_\rho)
-(1+\xi_s^2)(F_{1p}-\widetilde{F}_{1n}+F_{2p}-\widetilde{F}_{2n})G_\rho
\nonumber\\
&-&\biggl(\xi_s\xi_t\widetilde{F}_{1n}+2\eta\widetilde{F}_{2n}\biggr)
(G_\rho+F_\rho)\biggr\},\\
H_{kk}^{\mbox{w}}&=&H_{kk}^{\mbox{v}}+4Qm_{\scriptscriptstyle N}
\biggl(\frac{M_\rho^2-t}{m_{\scriptscriptstyle N}^2}\biggr)
\biggl(\xi_s\widetilde{F}_{1n}+\frac{Q}{2m_{\scriptscriptstyle N}}
\widetilde{F}_{2n})\biggr)F_\rho \,. 
\label{wkklv} \\
H_{ee}^{V}&=&4Q^2m_{\scriptscriptstyle N}^2
\biggl\{\xi_t^2\biggl[(F_{1p}-\widetilde{F}_{1n}+F_{2p}-\widetilde{F}_{2n})
\nonumber\\
&+&\frac{Q}{2m_{\scriptscriptstyle N}}
\biggl(\xi_s(F_{2p}+\widetilde{F}_{2n})-\xi_t\widetilde{F}_{2n}\biggr)
-\frac{Q^2}{4m_{\scriptscriptstyle N}^2}(F_{2p}+\widetilde{F}_{2n}) 
\biggr]G_\rho \nonumber\\
&-&4\eta\biggl[\biggl((1-\xi_s\xi_t+\xi_s^2)G_\rho-\eta{}F_\rho\biggr)
(F_{1p}-\widetilde{F}_{1n}+F_{2p}-\widetilde{F}_{2n}) \nonumber\\
&+&\frac{1}{2}\xi_t\biggl(\xi_s(F_{1p}+\widetilde{F}_{1n}
+F_{2p}+\widetilde{F}_{2n})-\xi_t(\widetilde{F}_{1n}+\widetilde{F}_{2n})
\biggr)(G_\rho+F_\rho)\nonumber\\
&-&\frac{Q}{2m_{\scriptscriptstyle N}}(\xi_s-\frac{1}{2}\xi_t)
\biggl((F_{1p}+\widetilde{F}_{1n})(G_\rho+F_\rho)
+(1+\xi_s^2)(F_{2p}+\widetilde{F}_{2n})
F_\rho-\xi_s\xi_t\widetilde{F}_{2n}F_\rho\biggr)\nonumber\\
&+&\frac{Q}{4m_{\scriptscriptstyle N}}(\xi_s\xi_t-2\eta)
\biggl(\xi_s(F_{2p}+\widetilde{F}_{2n})-\xi_t\widetilde{F}_{2n}\biggr)F_\rho
+\frac{Q}{4m_{\scriptscriptstyle N}}\xi_t(F_{2p}-\widetilde{F}_{2n})G_\rho
\nonumber\\
&-&\frac{Q^2}{4m_{\scriptscriptstyle N}^2}
(F_{2p}-\widetilde{F}_{2n})G_\rho+\biggl(\xi_s^2-\xi_s\xi_t+\eta\biggr)
(F_{2p}-\widetilde{F}_{2n})F_\rho
\biggr]\biggr\},\\
H_{ee}^{W}&=&H_{ee}^{V}+4Q^2m_{\scriptscriptstyle N}^2
\biggl(\frac{M_\rho^2-t}{m_{\scriptscriptstyle N}^2}\biggr)
\biggl\{\biggl(\frac{1}{2}\xi_s\xi_t-\frac{Q}{2m_{\scriptscriptstyle N}}
(\xi_s-\frac{1}{2}\xi_t)\biggr)
(F_{1p}+\widetilde{F}_{1n}+F_{2p}+\widetilde{F}_{2n})\nonumber\\
&+&\frac{Q^2}{4m_{\scriptscriptstyle N}^2}\xi_s(\xi_s-\xi_t)
(F_{2p}-\widetilde{F}_{2n})
-\frac{Q}{2m_{\scriptscriptstyle N}}\biggl(\xi_s(\xi_s-\xi_t)+\eta\biggr)
\biggl(\xi_s(F_{2p}+\widetilde{F}_{2n})-\xi_t\widetilde{F}_{2n}
\biggr)\nonumber\\
&-&\frac{1}{2}\xi_t^2(\widetilde{F}_{1n}+\widetilde{F}_{2n})
-\eta\biggl[(F_{1p}-\widetilde{F}_{1n}+F_{2p}-\widetilde{F}_{2n})
-\frac{Q^2}{4m_{\scriptscriptstyle N}^2}(F_{2p}-\widetilde{F}_{2n})
\biggr]\biggr\}F_\rho \,.
\label{weelv}
\en 

\subsection{Interference term for the ${\bm\rho}$-meson pole and the contact
${\bm\gamma}{\bm\pi}{\bf NN}$ vertex}
\eq 
& &\frac{1}{4\pi}\overline{\biggl[T_{\rho}^{V(W)}(\lambda)
T_{\scriptscriptstyle CPV}^\ast(\lambda)
+T_{\scriptscriptstyle CPV}
(\lambda)T_{\rho}^{V(W)\ast}(\lambda)\biggr]} = \nonumber\\
&=& \frac{\alpha}{4 m_{\scriptscriptstyle N}} F_\pi(Q^2) 
g_{\pi{\scriptscriptstyle NN}}(t) 
\frac{g_{\rho\pi\gamma}(t,Q^2)}{M_\rho^2-t}
\biggl\{(p\cdot\epsilon^{(\lambda)})(p\cdot\epsilon^{(\lambda)\,\ast})
Y_{pp}^{V(W)} 
+(k\cdot\epsilon^{(\lambda)})(k\cdot\epsilon^{(\lambda)\,\ast})
Y_{kk}^{V(W)}\nonumber\\
&+&\biggl[(p\cdot\epsilon^{(\lambda)})(k\cdot\epsilon^{(\lambda)\,\ast})
+(p\cdot\epsilon^{(\lambda)\,\ast})(k\cdot\epsilon^{(\lambda)})\biggr]
Y_{pk}^{V(W)}+(\epsilon^{(\lambda)}\cdot\epsilon^{(\lambda)\,\ast})
Y_{ee}^{V(W)}\biggr\} \,. \label{rpg}
\en 

\noindent
{\it Invariant factors} $Y_{\kappa\kappa^\prime}\,(\equiv
\Phi^{\rho{-} {\scriptscriptstyle CPV}}_{\kappa\kappa^\prime}(t,Q,W)\,)$:
\eq 
& &Y_{pp}^{V}=Y_{pp}^{W}=0\,, \qquad 
Y_{pk}^{V}=Y_{pk}^{W}=4m_{\scriptscriptstyle N}Q
\xi_t(G_\rho+F_\rho)\,,\qquad
Y_{kk}^{V}=Y_{kk}^{W}=-8m_{\scriptscriptstyle N}Q
\xi_s(G_\rho+F_\rho)\,,\nonumber\\
& &Y_{ee}^{V}=-32m_{\scriptscriptstyle N}^3Q\eta
\biggl(\xi_s-\frac{1}{2}\xi_t\biggr)
(G_\rho+F_\rho)\,,\qquad 
Y_{ee}^{W}=Y_{ee}^{V}+8m_{\scriptscriptstyle N}^3Q
\biggl(\frac{M_\rho^2-t}{m_{\scriptscriptstyle N}^2}\biggr)
\biggl(\xi_s-\frac{1}{2}\xi_t\biggr)F_\rho \,. 
\label{yvw}
\en 

\subsection{Interference term for the nucleon ($s+u$) poles and the contact
${\bm\pi}{\bm\gamma}{\bf NN}$ vertex}

\eq 
& &\frac{1}{4\pi}\overline{\biggl[T_{N(s{+}u)}(\lambda)
T_{\scriptscriptstyle CPV}^\ast(\lambda)
+T_{\scriptscriptstyle CPV}(\lambda)T_{N(s{+}u)}^\ast(\lambda)\biggr]}= 
\nonumber\\
&-&\alpha F_\pi(Q^2)\frac{g_{\pi{\scriptscriptstyle NN}}(t)}
{2m_{\scriptscriptstyle N}^2}\,
\frac{g_{\pi{\scriptscriptstyle NN}}F_{\rm eff}(t,Q^2)}
{W^2-m_{\scriptscriptstyle N}^2}\biggl\{(p\cdot\epsilon^{(\lambda)})
(p\cdot\epsilon^{(\lambda)\,\ast})X_{pp} 
+(k\cdot\epsilon^{(\lambda)})(k\cdot\epsilon^{(\lambda)\,\ast})X_{kk} 
\nonumber\\
&+&\biggl[(p\cdot\epsilon^{(\lambda)})(k\cdot\epsilon^{(\lambda)\,\ast})
+(p\cdot\epsilon^{(\lambda)\,\ast})(k\cdot\epsilon^{(\lambda)})
\biggr]X_{pk} 
+(\epsilon^{(\lambda)}\cdot\epsilon^{(\lambda)\,\ast})X_{ee}
\biggr\}\label{Npg}
\en 

\noindent
{\it Invariant factors} $X_{\kappa\kappa^\prime}\,(\equiv
\Phi^{N{-}{\scriptscriptstyle CPV}}_{\kappa\kappa^\prime}(t,Q,W)\,)$:
\eq 
X_{pp}&=&32m_{\scriptscriptstyle N}^2\biggl[
\frac{Q}{2m_{\scriptscriptstyle N}}\biggl(\xi_s(F_{1p}-\widetilde {F}_{1n})
+\xi_t\widetilde {F}_{1n}\biggr)
+\frac{Q}{4m_{\scriptscriptstyle N}}\xi_t(F_{2p}+\widetilde{F}_{2n}) 
-\frac{Q^2}{4m_{\scriptscriptstyle N}^2}
(F_{1p}+\widetilde{F}_{1n}+F_{2p}+\widetilde{F}_{2n})
\biggr]\,, \label{xpp}\\
X_{pk}&=&16m_{\scriptscriptstyle N}^2\biggl[
-\frac{Q}{2m_{\scriptscriptstyle N}}\biggl(\xi_s(F_{1p}-\widetilde {F}_{1n})
+\xi_t\widetilde {F}_{1n}\biggr)
-\frac{Q}{4m_{\scriptscriptstyle N}}\biggl(\xi_s(F_{2p}+\widetilde {F}_{2n})
+\xi_t\widetilde {F}_{2n}\biggr)\nonumber\\
&-&\frac{1}{2}(F_{1p}-\widetilde {F}_{1n})
+\frac{Q^2}{4m_{\scriptscriptstyle N}^2}
(F_{1p}+\widetilde{F}_{1n}+F_{2p}+\widetilde{F}_{2n})\biggr]\,,
\label{xpk}\\
X_{kk}&=&32m_{\scriptscriptstyle N}^2
\biggl(-\frac{1}{2}\widetilde {F}_{1n}
+\frac{Q}{4m_{\scriptscriptstyle N}}\xi_s\widetilde{F}_{2n}
\biggr)\,,\label{xkk}\\
X_{ee}&=&16m_{\scriptscriptstyle N}^3Q\biggl\{\biggl[
-\biggl(\eta\xi_s+\frac{1}{2}\xi_t\biggr)
+(1+\eta)\frac{Q}{2m_{\scriptscriptstyle N}}\biggr]
(F_{1p}+\widetilde{F}_{1n}+F_{2p}+\widetilde{F}_{2n})\nonumber\\
&+&\biggl(\xi_s-\frac{1}{2}\xi_t\biggr)
\biggl[2\eta(\widetilde {F}_{1n}+\widetilde{F}_{2n})
-\frac{Q^2}{4m_{\scriptscriptstyle N}^2}(F_{2p}-\widetilde{F}_{2n})
+\frac{Q}{2m_{\scriptscriptstyle N}}\biggl(\xi_s(F_{1p}+\widetilde {F}_{1n})
-\xi_t\widetilde {F}_{1n}\biggr)\biggr]
\biggr\} \,.\label{xee}
\en 
\vspace*{3cm} 
\end{widetext}

\newpage

\begin{widetext}

\begin{center}
{\bf Table 1.} Comparison of results for the pion form factor $F_\pi(Q^2)$ 

\vspace*{.3cm} 

\def\arraystretch{1.2}
\begin{tabular}{c|cccc|cc}
& \multicolumn{4}{|c|}{F$\pi$1 data~\cite{Tadevosyan:2007yd}} &
{F$\pi$2 data~\cite{Horn:2006tm}}\\[3pt]
\hline
$Q^2$ (GeV$^2$/c$^2$) & 0.6 & 0.75 & 1 & 1.6 & 1.6 & 2.45 \\[3pt]  
\hline
$F_\pi$~\cite{Horn:2006tm,Tadevosyan:2007yd} & 0.433 & 0.341 & 0.312 
& 0.233 & 0.243 & 0.167 \\[3pt]
& $\pm$ 0.017 & $\pm$ 0.022 & $\pm$ 0.016 & $\pm$ 0.014 & $\pm$ 0.012 
& $\pm$ 0.010 \\[3pt]
\hline
$F_\pi(V)$ & 0.412 & 0.348 & 0.309 & 0.239 & 0.242 & 0.168 \\[3pt]
$\Lambda_\pi^2(V)$ & 0.420 & 0.400 & 0.447 & 0.503 & 0.511 & 0.494 \\[3pt]
\hline
$F_\pi(W)$ & 0.420 & 0.368 & 0.335 & 0.294 & 0.272 & 0.200 \\[3pt]
$\Lambda_\pi^2(W)$ & 0.434 & 0.437 & 0.504 & 0.666 & 0.597 & 0.614 \\[3pt]
\end{tabular}
\end{center}

\vspace{.5cm}

\begin{center}

{\bf Table 2.} Coupling constants and cutoff parameters 

\vspace*{.2cm}

\def\arraystretch{1.2}
\begin{tabular}{c|c|c|c|c|c|c}
$g_{\pi{\scriptscriptstyle NN}}$ & $g_{\rho\pi\gamma}$ &
$G_{\rho{\scriptscriptstyle NN}}(0)$ & $F_{\rho{\scriptscriptstyle NN}}(0)$ &
$Z$ & $\Lambda_{\rm str}^2$ & $\Lambda_{\rm eff}^2$\\[3pt]
& GeV$^{-1}$ &&&& GeV$^2$/c$^2$ & GeV$^2$/c$^2$\\[3pt]
\hline
13.5 & 0.728 & 4 & 26 & 0.11 & 0.7 & 1.44\\
\end{tabular}
\end{center}

\vspace{.5cm}

\begin{center}
{\bf Table 3.} Predicted values of $\Lambda_\pi^2$ for a variety of
theoretical approaches 

\vspace*{.3cm} 

\def\arraystretch{1.2}
\begin{tabular}{c|l}
$\Lambda_\pi^2$ (GeV$^2$/c$^2$)& $\qquad$ Theory \\[3pt]
\hline
0.51& Extended Nambu-Jona-Lasinio Model~\cite{Anikin:1995cf}\\[3pt]
0.52& Bethe-Salpeter/Schwinger-Dyson Equations~\cite{Maris:2000sk}\\[3pt]
0.54& Light Front Dynamics~\cite{Cardarelli:1995hn}\\[3pt]
0.55& Relativistic Quark Model~\cite{Faessler:2003yf}\\[3pt]
0.60& Nonlocal Chiral Quark Model~\cite{Dorokhov:2003sc}\\[3pt]
0.66& Bethe-Salpeter/Schwinger-Dyson Equations~\cite{Roberts:1994hh}\\[3pt]
0.66& QCD Sum Rules~\cite{Nesterenko:1982gc}\\
\end{tabular}
\end{center}

\newpage

%%%%%%%%%%%%%%%%%%% Fig. 1 %%%%%%%%%%%%%%%%%%%%%%%%%%%%%%%
\begin{figure}[hp]
\begin{center}

\hspace*{7cm}
\epsfig{file=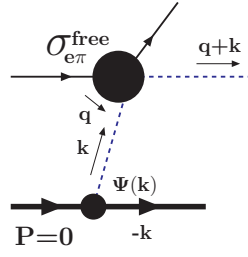,scale=.7}
\end{center}

\vspace*{.5cm}
 
\caption{(color online). 
Dominating $t$-channel quasi-elastic mechanism 
for the longitudinal part $\sigma_L$ of the cross section.}  
\label{fig1}
\end{figure}
%%%%%%%%%%%%%%%%%%%%%%%%%%%%%%%%%%%%%%%%%%%%%%%%%%%%%%%%%%

\vspace*{.25cm} 

%%%%%%%%%%%%%%%%%%%%%%% Fig. 2 %%%%%%%%%%%%%%%%%%%%%%%%%%%%%%%%%%%%%%
\begin{figure*}[hp]
\begin{center}
\epsfig{file=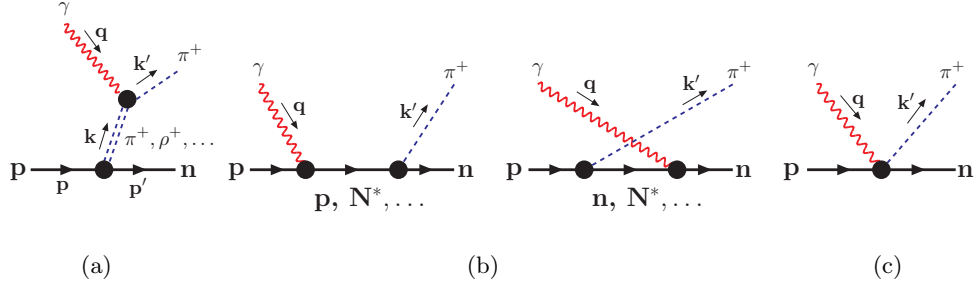,scale=.7}
\end{center}

\hspace*{.75cm}
(a) \hspace*{4.5cm} 
(b) \hspace*{4.75cm} (c)

\caption{(color online). 
Feynman diagrams (``Born approximation'') 
for the pion electroproduction off the nucleon.}

\label{fig2}

\end{figure*}
%%%%%%%%%%%%%%%%%%%%%%%%%%%%%%%%%%%%%%%%%%%%%%%%%%%%%%%%%%%%%%%%%%%%%

\vspace*{.5cm} 

%%%%%%%%%%%%%%%%%%% Fig. 3 %%%%%%%%%%%%%%%%%%%%%%%%%%%%%%%
\begin{figure}[hp]
\begin{center}
\epsfig{file=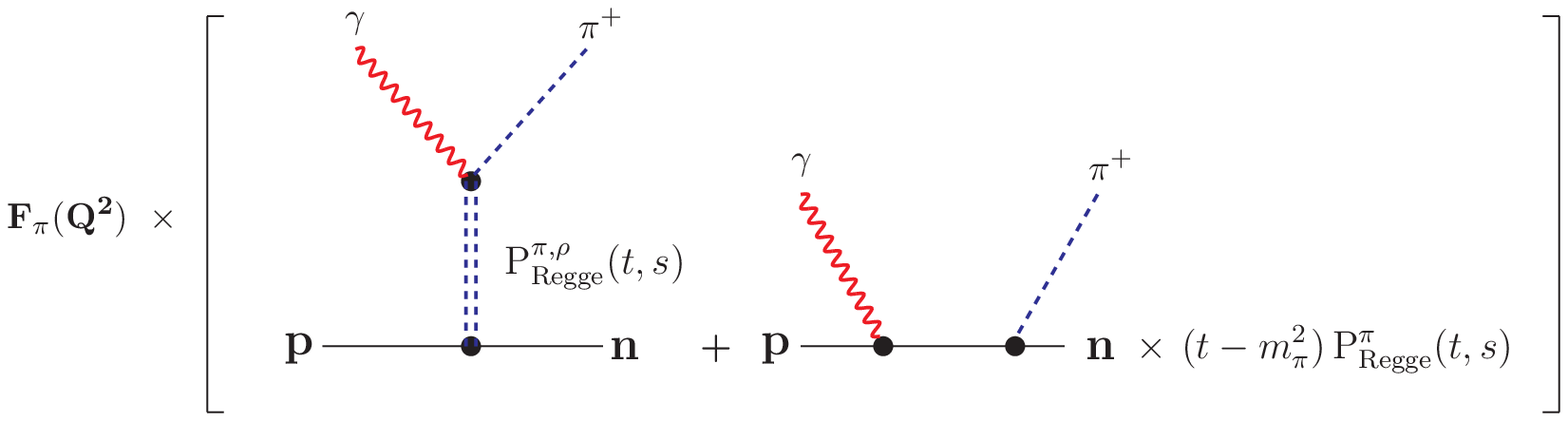,scale=.7}

\vspace{.25cm}

\caption{(color online). 
Factorization of the amplitude of pion electroproduction 
off the nucleon in the Regge model~\cite{Vanderhaeghen:1997ts}.} 
\label{fig3}
\end{center}
%%%%%%%%%%%%%%%%%%%%%%%%%%%%%%%%%%%%%%%%%%%%%%%%%%%%%%%%%%

%%%%%%%%%%%%%%%%%%% Fig. 4 %%%%%%%%%%%%%%%%%%%%%%%%%%%%%%%
\begin{center}

\vspace*{2.5cm} 

\epsfig{file=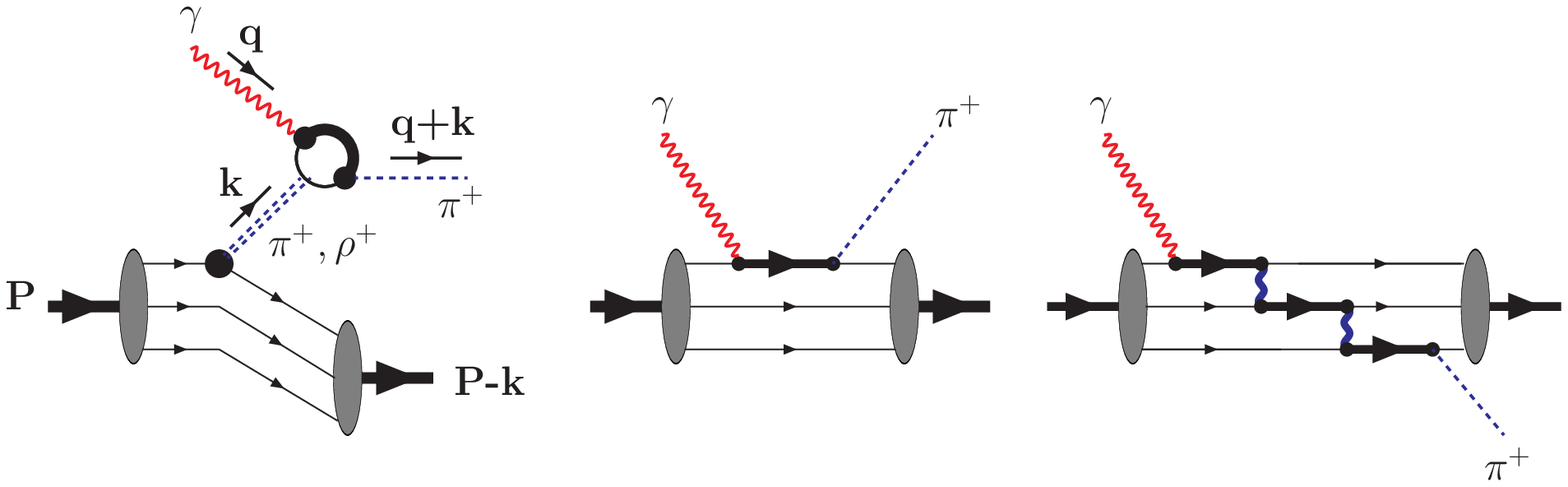,scale=.7}
\end{center}

\vspace*{-1cm}

(a) \hspace*{4.5cm} 
(b) \hspace*{4cm} (c) 

\caption{(color online). 
Microscopic interpretation of $t$- and
$s$-channel processes within the quark model. Thick lines indicate
the propagation of a large photon momentum $q$ through separate
partons inside the nucleon.}

\label{fig4}

\end{figure} 
%%%%%%%%%%%%%%%%%%%%%%%%%%%%%%%%%%%%%%%%%%%%%%%%%%%%%%%%%%

\newpage 

%%%%%%%%%%%%%%%%%%% Fig. 5 %%%%%%%%%%%%%%%%%%%%%%%%%%%%%%%
\begin{figure}
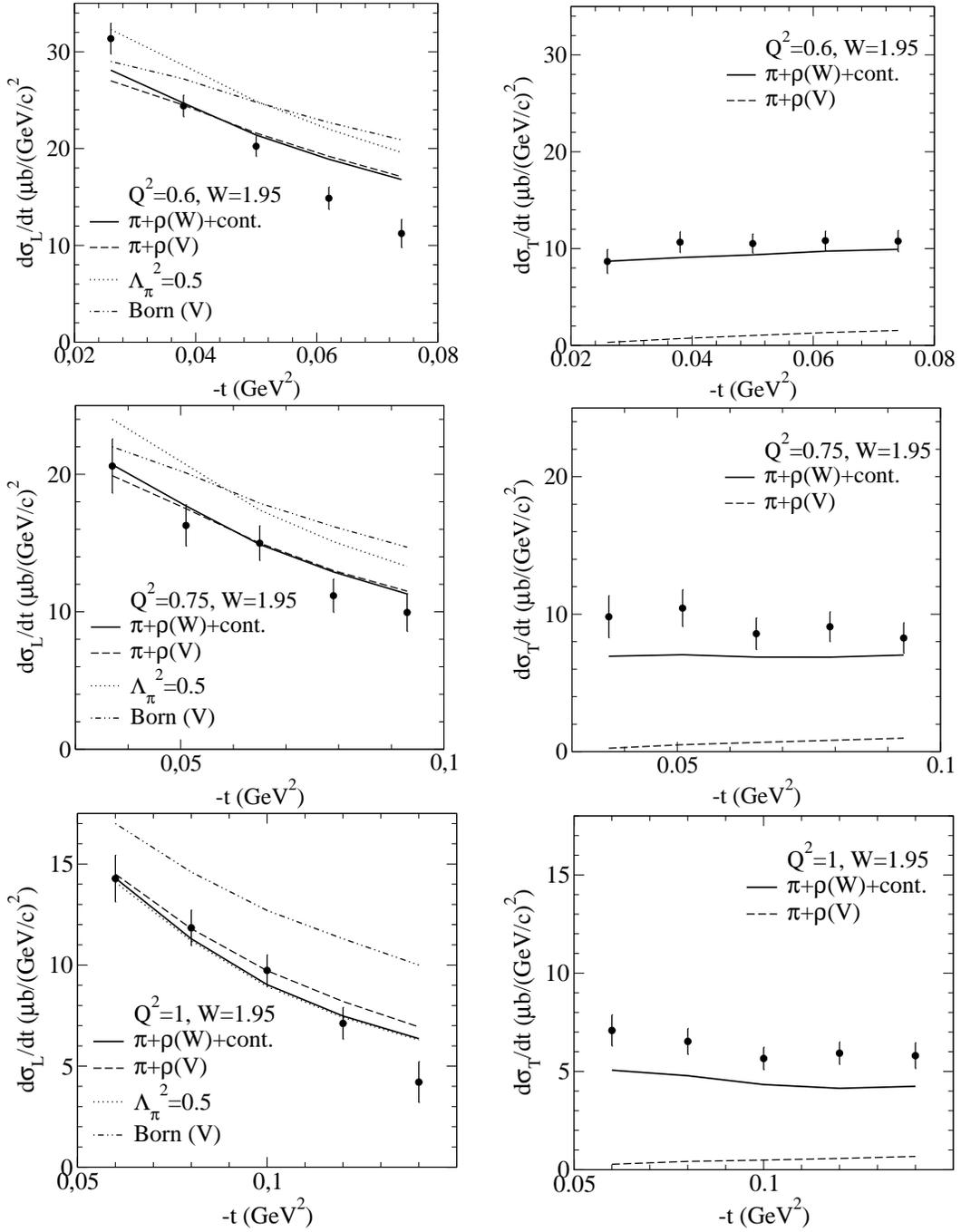

\begin{center}
\mbox{
{\epsfig{figure=ffL06a.eps,width=0.37\textwidth,clip}}\qquad
{\epsfig{figure=ffT06.eps,width=0.37\textwidth,clip}}
\vspace{0.3cm}
}
\mbox{
{\epsfig{figure=ffL075a.eps,width=0.37\textwidth,clip}}\qquad
{\epsfig{figure=ffT075.eps,width=0.37\textwidth,clip}}
\vspace{0.3cm}
}
\mbox{
{\epsfig{figure=ffL10a.eps,width=0.37\textwidth,clip}}\qquad
{\epsfig{figure=ffT10.eps,width=0.37\textwidth,clip}}
}
\end{center}
\caption{Longitudinal and transverse cross sections for
the $p(e,e^{\prime}\pi^+)n$ reaction at small $Q^2\lesssim$ 1  
GeV$^2$. The unscaled $F\pi1$ 
data are centered at $Q^2$ of 0.6,  0.75 and 1.GeV$^2$.  
Interference terms between $\pi$-pole, $\rho$-pole and contact
$\pi\gamma NN$ amplitudes are taken into account. 
Results obtained in the $t$-pole
approximation with the standard vector (V) $\rho$-meson field
are shown by dashed lines.
Results for the tensor variant (W) of the $\rho$-meson field 
(the contact diagram is included) are shown by solid lines
(by dotted lines for the fixed value of $\Lambda_\pi^2=$ 0.5 GeV$^2$/c$^2$).
Note that the proper values of $W$ and $Q^2$ for each -$t$ bin
are different and they differ from the average values shown in the figure 
legends (see \cite{Horn:2006tm,Tadevosyan:2007yd} for detail).
For comparison the proper Born approximation results (i.e. without the strong
vertex form factors) are shown by double-dot dashed lines.}
\label{fig5}
\end{figure}
%%%%%%%%%%%%%%%%%%%%%%%%%%%%%%%%%%%%%%%%%%%%%%%%%%%%%%%%%%

\newpage 

%%%%%%%%%%%%%%%%%%% Fig. 6 %%%%%%%%%%%%%%%%%%%%%%%%%%%%%%%
\begin{figure}
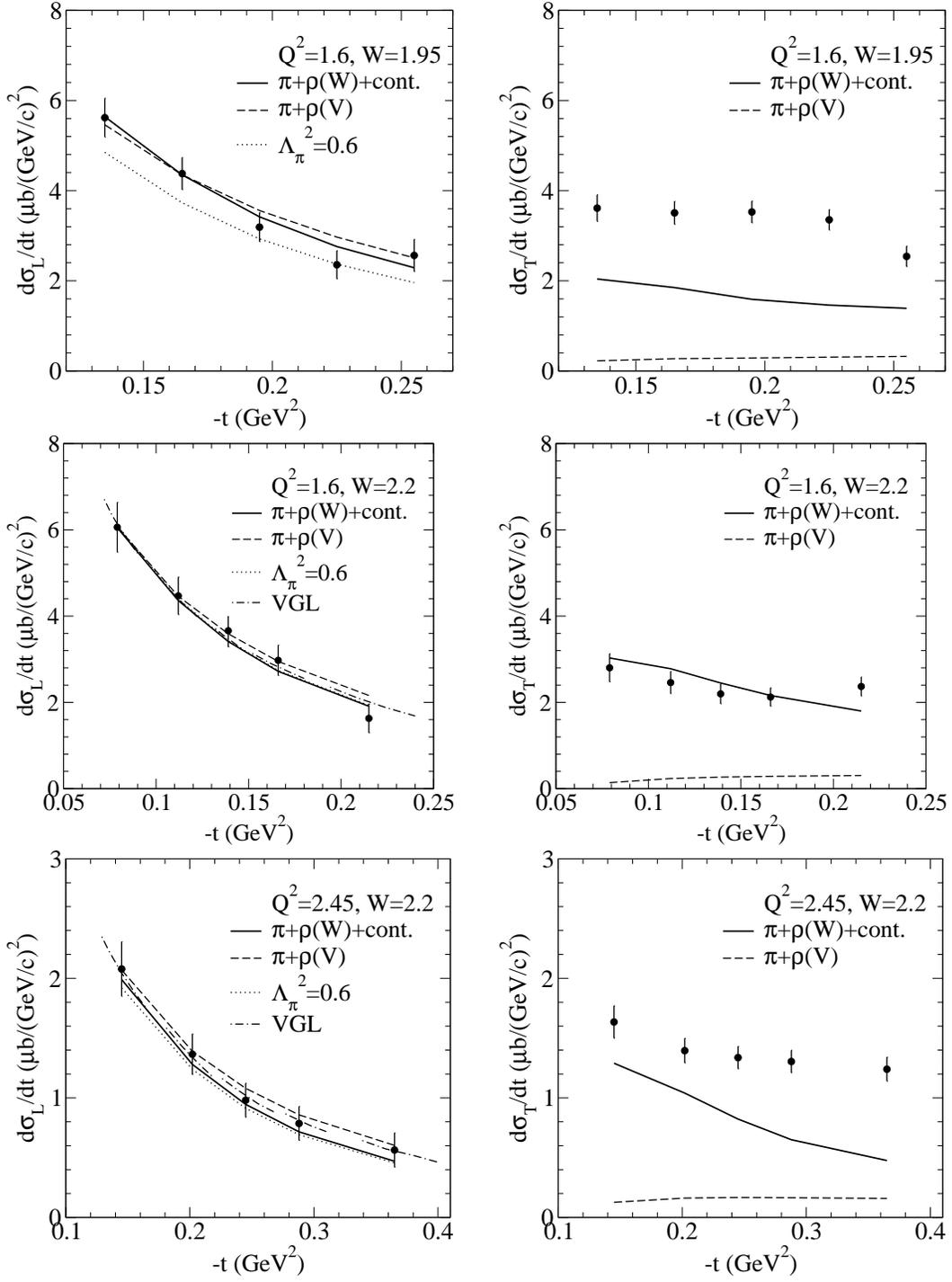

\begin{center}
\mbox{
{\epsfig{figure=ffL16.eps,width=0.37\textwidth,clip}}\qquad
{\epsfig{figure=ffT16.eps,width=0.37\textwidth,clip}}
\vspace{0.3cm}
}
\mbox{
{\epsfig{figure=ffL2_16.eps,width=0.37\textwidth,clip}}\qquad
{\epsfig{figure=ffT2_16.eps,width=0.37\textwidth,clip}}
\vspace{0.3cm}
}
\mbox{
{\epsfig{figure=ffL2_25.eps,width=0.37\textwidth,clip}}\qquad
{\epsfig{figure=ffT2_25.eps,width=0.37\textwidth,clip}}
}
\end{center}
\caption{Longitudinal and transverse cross sections at larger values of $Q^2$: 
1.6\,GeV$^2$/c$^2$ for the $F\pi$1 data and 1.6 and 2.45\,GeV$^2$/c$^2$
for the $F\pi$2 data. The same notation as in Fig.~\ref{fig5}. 
The dotted lines correspond to the fixed value of 
$\Lambda_\pi^2=$ 0.6 GeV$^2$/c$^2$.
Here for comparison the VGL model results (adapted from 
Ref.~\cite{Horn:2006tm}) are shown by dash-dotted lines.}
\label{fig6}
\end{figure}
\end{widetext}

%%%%%%%%%%%%%%%%%%%%%%%%%%%%%%%%%%%%%%%%%%%%%%%%%%%%%%%%%%

\end{document}